\documentclass[12pt]{article}
\usepackage[english]{babel}
\usepackage{amsmath,epsf}
\usepackage[T1]{fontenc}
\usepackage[latin1]{inputenc}
\usepackage{amsfonts,amssymb}
\usepackage{yfonts}
\usepackage[usenames,dvipsnames]{color}
\newtheorem{thm}{Theorem}

\newtheorem{lemma}{Lemma}[section]
\newtheorem{prop}{Proposition}[section]

\newcommand{\al}{\alpha}

\newcommand{\be}{\beta}
\newcommand{\de}{\delta}

\newcommand{\vep}{\varepsilon}
\newcommand{\ga}{\gamma}
\newcommand{\Ga}{\Gamma}

\newcommand{\la}{\lambda}

\newcommand{\vp}{\varphi}
\newcommand{\La}{\Lambda}

\newcommand{\Lao}{\Lambda_0}

\newcommand{\pa}{\partial}

\newcommand{\ti}[1]{\tilde{#1}}

\newcommand{\qed}{\hfill \rule {1ex}{1ex}\\ }

\newcommand{\eq}{\begin{equation}}
\newcommand{\eqe}{\end{equation}}
\newcommand{\ben}{\begin{equation}}
\newcommand{\een}{\end{equation}}

\newcounter{saveeqn}

\begin{document}

\title{Mean field flow equations and asymptotically
free scalar fields}
\author{Christoph Kopper$^{1}$
\thanks{\tt christoph.kopper@polytechnique.edu}\:
\\ \\
{\it ${}^{1}$Centre de Physique Th\'eorique CPHT, CNRS, UMR 7644} \\
{\it  Institut Polytechnique de Paris, 91128 Palaiseau, France} 
}  

\maketitle
\begin{abstract}
The flow equations of the renormalisation group
permit to analyse the perturbative $n$-point functions of 
renormalisable quantum field theories. Rigorous bounds  
implying renormalisablility allow  to control large 
momentum behaviour, infrared singularities and large order
behaviour in the number of loops and the number of arguments 
$n\,$. Gauge symmetry which is broken by the flow in momentum
or position space, can be shown to be restored in the 
renormalised theory. 

In this paper we want to do a first but important step
towards a rigorous nonperturbative analysis of the flow equations  
(FEs). We restrict to massive scalar fields and analyse the 
{\it mean field limit} where the Schwinger or 1PI functions 
are considered to be momentum independent or, otherwise stated, 
are replaced  by their zero momentum values. 
 We regard smooth solutions of the system of FEs 
for the $n$-point functions for different 
sets of boundary conditions. We  will realise 
that allowing for non\-vanishing irrelevant terms 
permits to construct {\it asymptotically free} and thus
nontrivial {\it scalar field theories} in the mean field 
approximation.  We will also analyse  the so-called trivial
solution so far generally believed to exhaust 
four-dimensional scalar field theory. The method paves 
the way to a study of 
the system of FEs beyond the mean field  limit.
\end{abstract}


\section{Introduction}

Quantum field theory, originally developed to implement the
principles of quantum mechanics in relativistic systems, 
has become the general theoretical framework to study
physical systems with an infinite (or large) number of degrees
of freedom. Relativistic quantum systems are described by relativistic
quantum field theory, Euclidean field theory gives access to critical systems
in statistical mechanics, systems from solid state physics 
can be modeled by field theories at finite density and temperature.
These systems have different kinematics which is reflected 
in particular by the form of the (free) propagator or two-point function.
Interactions are introduced via the  path integral
formalism.
   
Aiming  at mathematical rigour one
is faced with the problem that path integrals describing interacting
systems in field theory are generally not defined a priori. 
Whereas there is a complete theory of Gaussian measures applying to
the noninteracting case, a mathematically oriented study of 
interacting field theories generally starts from regularised
versions of the theory, where the number of degrees of freedom
in space and momentum has been made (essentially) finite by hand, 
through the introduction of regulators like finite volume and
large momentum cutoffs. One then studies correlation functions 
and proves that these have uniform limits in the cutoffs. 
For a general introduction to these methods 
see ~\cite{Glimm}, euclidean scalar field theories
are analysed in ~\cite{Riv:PCR, Simon}. 

The functional flow equation is a differential equation for the effective
action functional of the field theory considered. When
expanded in moments it becomes an infinite system 
of differential equations for the connected
amputated Schwinger functions of the theory. 
In a seminal paper ~\cite{Polchinski:1983gv} Polchinski observed that
when expanding these functions order by order in the number of loops, 
there is an airtight inductive scheme which permits 
to sufficiently control the perturbative functions such that 
renormalisability follows. In a subsequent
paper ~\cite{Keller:1990ej}, see also ~\cite{Keller:1991bz},
and ~\cite{Kop, Muller:2002he} for reviews,   
it was shown how to impose
physical renormalisation conditions, and the induction
hypothesis was sharpened so that cutoff independence
became immediate. As a result  ultraviolet 
renormalisability could be largely reduced to power 
counting once the optimal induction hypothesis had been
found. The complicated combinatoric aspects of the
problem, which had found their deep solution in Zimmermann's
forest formula \cite{Zi69}, thus turned out not to be
intrinsic to the renormalisation problem, but rather
to stem from the fact that the perturbative contributions
had been split up in too fine a way, namely into Feynman
diagram amplitudes\,\footnote
{This remark does of course not put into question  
the value of Feynman diagrams. 
It only says that they are not optimally adapted for a 
mathematical analysis of the UV divergences (and related 
problems).
}.
In contrast, methods originally stemming from statistical 
physics like cluster and Mayer expansions ~\cite{MayerMontroll, Brydges, Glimm,
Riv:PCR, BBS}, 
permit to analyse regularised path integrals nonperturbatively,
but are relatively straightforward to apply only in theories which do not 
have to be renormalised in an essential way like 
$\vp^4_2\,$ and $\vp^4_3\,$ or other superrenormalisable
models. They are technically very hard to apply in strictly
renormalisable theories. The review ~\cite{BBS}  
shows the state of the art and reveals important progress
made in this respect over the last decades. 
Still in quantum field theory with hindsight to particle physics
the relevance of the 
constructive path integral method is also limited 
by the fact that the physically interesting theories
are either plagued by the triviality statement or 
by infrared problems which are presently beyond scope
in mathematical physics, as is the case for
quantum chromodynamics. 
As a consequence the work
performed in constructive field theory
has not entered text books on quantum field theory
outside the realm of mathematical physics, 
in spite of the fact that the nonperturbative 
analysis of field theory is generally recognized to be 
an important problem. 
               
Regarding on the other hand the flow equations,
one realises that they  have not been used 
with much success in the rigorous analysis of quantum 
field theory beyond perturbation theory. Whereas
the renormalisation problem becomes transparent and easy
in this framework, by being related immediately to 
power counting,
there are problems of combinatoric or algebraic
origin, which hinder a nonperturbative analysis,
even in the absence of renormalisation. 
To some degree these problems are already present in $\vp^4_1\,$ 
or even  $\vp^4_0\,$ theory as will be explained 
more precisely in the following. In this
context we cite a beautiful paper by 
Rivasseau ~\cite{Riv84} on a construction of planar 
``wrong sign'' $\vp^4_4\,$-theory  which is intermediate 
between constructive and flow equation methods.

In this paper we shall consider scalar field theories
in four dimensions
in the mean field limit. These have
the same power counting and scaling behaviour 
as the full four dimensional theory and seem to 
capture well the basic physical properties
of four dimensional scalar field theories.   

In perturbative quantum  field
theory, one typically starts from a bare action
which contains only a few local monomials of low degree
in the fields. In a theory like quantum electrodynamics
this leads, after perturbative renormalisation, to 
results which are in extremely good agreement with 
experiment. The simple form of the bare action
is also justified by the fact that 
higher order monomials in the fields lead to 
nonrenormalisable interactions. Theories containing 
such interactions are generally not 
predictive in the perturbative framework. 
From the point of view of the Wilson renormalisation 
group these low order monomial lagrangians correspond
to a  fine tuning procedure.  
Staying with the example of quantum electrodynamics, 
this means that even if we start at the UV cutoff scale 
with a bare lagrangian containing only the monomials 
\[
\bar \psi\, \psi \ ,\quad \bar \psi\,\pa^\mu \psi 
\ ,\quad \bar \psi\, A^\mu\, \psi \ ,
\quad F_{\mu\nu}F^{\mu \nu}\ , \quad (\pa_\mu A^\mu)^2 \ ,   
\]
after integrating out degrees of freedom (in whatever small 
a momentum range) we immediately obtain 
a nonpolynomial effective lagrangian containing 
monomials of any degree in the fields and their derivatives
(as far as they are allowed by the symmetries of the theory).
Otherwise stated, this fine tuning consists in arranging things 
such that the infinite number of trajectories of all
higher n-point functions are forced to pass through $0$ each,
and for all momentum arguments,
at the same value of the renormalisation scale,
leaving only the few local terms appearing in the bare lagrangian
we wrote above. 
 
The infinite number of terms generated by the renormalisation 
group evolution can be shown in perturbation theory to contain 
each an inverse power of the renormalisation group scale, 
corresponding to its mass dimension, times 
a suitable function bounded uniformly in the scale up to 
logarithms. All these terms are uniformly bounded in the UV cutoff 
~\cite{KKS, Muller:2002he}. Their contributions do not produce any 
new ultraviolet
divergences, as compared to those stemming from the initial bare 
lagrangian. Thus the argument of nonrenormalisability disappears
in this case due to the aforementioned inverse powers.  
Nevertheless one also tends to start from those low order 
monomial bare lagrangians in constructive field theory which is 
based on the Wilson renormalisation group. There are two reasons for 
this: the starting point of the construction is easier to
control mathematically, and perturbative calculations in physics
are based on monomial bare lagrangians.
Nor does it seem evident to characterize generic 
classes of nonpolynomial starting lagrangians.

In this paper we want to insist on the fact that different,  
generally nonpolynomial, bare  actions, scaling with the cutoff
as indicated  before, may lead to
essentially different theories. 
As we said we restrict to four 
dimensional scalar field
theories in the mean field limit. We will show that, depending 
on the choice of the bare lagrangian, one may in particular 
obtain asymptotically 
free scalar field theories which escape the so-called triviality
statement ~\cite{Froehlich82, Aizenman82}. Our results 
are not in contradiction with this statement
since we will verify that for the fine tuned bare Lagrangian
containing only local terms of the type $\vp^4$ and $\vp^2$,
 the trivial solution emerges indeed.
 We will characterize this solution quite explicitly. 
It will turn out  that enforcing the fine tuned boundary 
conditions generates large values for the derivatives of the n-point functions 
w.r.t. the renormalisation group scale.  

We think our results are robust, in the sense 
that we expect them to hold beyond the mean field limit.
This is due in particular to the fact that we find 
the trivial solution as expected. From the technical 
side the main point to be mentioned is that when going 
beyond the mean field limit, there appears
a function $\,\dot C^\al(p)\,$ in the FEs, see \eqref{fe},
which is replaced by $\,1\,$ in the limit. Generally 
we have $\,0\,< \, \dot C^\al(p)\,=\,e^{-\al(p^2+m^2)}\,\le \,1\,$, 
which means that 
taking into account this factor leads to a contraction of 
the respective term in FEs which should be a controllable
modification.
We also mention that the critical behaviour 
in statistical physics is exactly described by the mean
field approximation in $\,d>4\,$ dimensions~\cite{Froehlich82, Aizenman82},
as was first pointed out by Ginzburg~\cite{Ginzburg}.

The conclusion of our findings is that irrelevant terms
in the language of the renormalisation group, 
can nevertheless completely modify the behaviour
of a theory, at least when we add an infinity
of such terms,

The paper is organized as follows\,:
In section \ref{sec:fe} we introduce the flow equations.
In section \ref{sec:meanfield} we perfom the mean field
limit. Section \ref{sec:solutions} is at the heart of this
paper. We study various types of smooth solutions of 
mean field FEs.  In \ref{ssec:solundet} we study solutions
for which we impose certain smallness and smoothnees conditions 
at the UV boundary. Then in \ref{ssec:asy+} we study a class
of strictly
positive solutions of similar type.
In \ref{ssec:solbounded} we also impose bounds on the initial
conditions which are sufficient to make the starting 
regularised path integral well defined. 
In \ref{ssec:trivialsol} we study the trivial solution.
In section \ref{sec:1PI} we develop the corresponding
formalism for one-particle irreducible functions
and obtain results corresponding to section \ref{ssec:solundet}
for those.


\section{The flow equations}
\label{sec:fe}

We consider a self-interacting scalar field  on four dimensional 
Euclidean space. We adopt the renormalisation group flow equation 
framework~\cite{Wilson:1971bg,Wilson:1971dh,Wegner:1972ih,Polchinski:1983gv}. 
In the following  we will give a brief review of the general formalism and 
define the  objects of interest for the purpose of this paper. 
See~\cite{Muller:2002he, Kop, Hollands:2011gf} for more 
comprehensive reviews of the flow equation approach within our context.


\subsection{The flow equations for the effective action}
\label{ssec:feea}

We start formulating our theory with ultraviolet (UV) cutoff 
 and infrared (IR) cutoff in the standard path integral 
formalism. This requires two main ingredients:
\begin{enumerate}
\item We define the regularised momentum space propagator as
\ben
C^{\al_{0},\al}(p;m)=\frac{1}{p^{2}+m^2}\, \left[ \exp\left(-\al_0(p^{2}+m^2)\right)
-\exp\left(-\al (p^{2}+m^2)\right) \right]\ .
\label{cov}
\een
Upon removal of the cutoffs, i.e. in the limit 
$\al_0\to 0$ (UV), $\al\to \infty$ (IR), we indeed recover the free  propagator 
$\frac{1}{p^{2}+m^2}\,$.
For the Fourier transform, we use the convention
\ben
f(x) = \int_p  \hat f(p)\, e^{ipx} := \int_{\mathbb{R}^4} \frac{d^4 p}{(2\pi)^4}\, 
\,e^{ipx} \hat f(p) 
\een
so that in position space
\eq
{C}^{\al_0,\al}(x-y;m) = 
\int_p  e^{ip(x-y)}\ C^{\al_0,\al}(p;m) \ \ 
 \mbox{ using the shorthand }\ \
\int_p  := \int \frac{d^4p}{(2\pi)^4}\ .
\eqe
\item The interaction Lagrangian is supposed to be of 
the form
\footnote
{Since we will pass to the mean field limit
soon, we do not introduce a momentum dependent wave function
renormalisation term $b_0 \, (\partial \varphi(x))^2\,$ here.
}
\ben
   L_{0}(\vp) =  \int  d^4x\ 
\sum_{n \in 2\mathbb{N}} c_{0,n}(\al_0)\  \vp^{n}(x)\ ,
\label{bare}
\een
where the constants  $\,c_{0,n}\,$ should be such that 
\ben
- \infty \  <\  K\ <\  L_{0}(\vp)\  <\  \infty
\quad \ \forall \ \vp \in {\rm supp}\ \mu^{\al,\al_0} 
\label{positivity}
\een
for some finite real constant $\,K\,$.
The \emph{basic field} $\varphi $ is assumed to be in the 
support  of the normalised Gaussian measure $\mu^{\al,\al_0}$
 with covariance
\eqref{cov}
\footnote{
See the Appendix to Part I of~\cite{Glimm} for mathematical details 
about Gaussian functional integrals.}. 
In order to obtain a well defined limit of the quantities of interest, 
the constants $c_{0,n}$ generally need to be chosen as appropriate functions of 
the ultraviolet cutoff $\al_0$.
\end{enumerate}
The correlation ($=$ Schwinger $=$ $n$-point) functions of 
$n$ basic fields with cutoff are defined by the expectation values
\ben\label{pathint}
\begin{split}
 \langle \varphi(x_1) \cdots \varphi(x_n) \rangle &\equiv  
\mathbb{E}_{\mu^{\al,\al_0}} \bigg[\exp \bigg( -
L_0\bigg) \, \varphi(x_1) \cdots \varphi(x_n) \bigg] \bigg/ Z^{\al,\al_0} \\
& =
\int d\mu^{\al,\al_0} \ \exp \bigg( -
L_0\bigg) \, \varphi(x_1) \cdots \varphi(x_n)\bigg/ Z^{\al,\al_0} \, .
\end{split}
\een
This expression is simply the standard Euclidean path-integral, but with 
the free part in the Lagrangian absorbed into
$\,d\mu^{\al,\al_0}\,$.
The normalisation factor $Z^{\al,\al_0}$
is chosen so that $\langle 1 \rangle = 1$. 
For finite values of the cutoffs $0<\al_0< \al <\infty\, $ 
and on imposing a finite (space) volume, the functional integral 
\eqref{pathint} exists in the nonperturbative sense. 
In the perturbative theory it has been  shown that one can remove the 
cutoffs, $\al_0 \to 0$ and $\al \to \infty$, for a suitable choice 
of the running couplings $c_{0,n}(\al_0)$ at each given but fixed order in the 
number of loops. 
The correct behaviour of these couplings (in terms of bounds) is determined 
from the FEs which are  a system of differential equations in the parameter
$\al$ for the Schwinger functions.

These differential equations are written most conveniently in
terms of the hierarchy of ``connected, amputated
Schwinger functions'' (CAS functions),
whose generating functional is given by the following 
convolution\footnote{The convolution is defined in general by
$(\mu^{\al_0,\al} \star F)(\varphi) =
\int d\mu^{\al_0,\al}(\varphi') \ F(\varphi+\varphi')$.}
of the Gaussian measure with the exponentiated interaction,
\ben\label{CAGdef}
-L^{\al_0,\al} := \, \log \left[ \mu^{\al_0,\al}
\star \exp \bigg(- L_0 \bigg)\right] -  \log  Z^{\al_0,\al} \ .
\een
The full Schwinger functions can be recovered from the CAS functions
 in the end. 
One can expand the functionals $L^{\al_0,\al}$ as formal power series 
in terms of Feynman diagrams with $\ell$ loops, $n$ external legs and 
propagator $C^{\al_0,\al}(p)$. One can show that, indeed, only connected 
diagrams with an even number of external legs
contribute, and that the (free) propagators on the external legs 
are removed. While we will not use diagrammatic decompositions in terms 
of Feynman diagrams, we start from analyzing the functional (\ref{CAGdef})
in momentum space, expanded in moments, i.e. powers of $\varphi$
\ben\label{genfunc}
L^{\al_0,\al}(\varphi) := \sum_{n\in 2 \mathbb{N}}
\int \frac{d^4 p_1}{(2\pi)^{4}} \dots 
\frac{d^4 p_n}{(2\pi)^{4}}\ \bar{\cal L}^{\al_0,\al}_{n}(p_1, \dots, p_n)
\,
\hat\varphi(p_1) \cdots \hat\varphi(p_n) \, .
\een
Here no statement is made about the
convergence of this series. 
By performing the Fourier transformation in \eqref{bare} we find the relation
\ben
(2\pi)^4\; c_{0,n}(\al_0)\ \de^{4}(\sum_{i=1}^n p_i)
\,=\,\bar{\cal L}^{\al_0,\al_0}_{n}(p_1, \dots, p_n)\ .   
\label{c0}
\een
Translation invariance of the CAS functions in 
position space implies that
$\bar{{\cal L}}^{\al_0,\al}_{n}(p_1, \dots, p_n)$ is supported
at $p_1+\ldots+p_n=0$ 
(momentum conservation), and thus only depends on $n-1$ independent four momenta. 
We write
\ben
\label{deltaCAG}
\bar{{\cal L}}^{\al_0,\al}_{n}(p_1, \dots, p_n)\,= \,
\delta^{4}{(\sum_{i=1}^n p_i)}\ {\cal L}^{\al_0,\al}_{n}(p_1, \dots, p_n) 
\een
so that 
\ben
c_{0,n}(\al_0)\,=\,  (2\pi)^{-4}\,{\cal L}^{\al_0,\al_0}_{n}(p_1, \dots, p_n)\ .
\label{con}
\een 
We use the convention that the variable $p_{n}$ 
is determined in terms of the remaining $n-1$
 four vectors by momentum conservation, i.e. $p_{n}=-p_{1}-\ldots -p_{n-1}$.
  One should keep in mind, however, that the functions 
$\bar{{\cal L}}^{\al_0,\al}_{n}(p_1, \dots, p_n)\,$ are in fact fully 
symmetric under permutation of $p_{1},\ldots, p_{n}$.

To obtain the flow equations for the CAS functions, 
we take the $\al$-derivative of
\eqref{CAGdef}:
\ben
\partial_{\al} L^{\al_0,\al} \,=\,
\frac{1}{2}\,
\langle\frac{\delta}{\delta \vp},\,\dot {C}^{\al}\star
\frac{\delta}{\delta \vp}\rangle L^{\al_0,\al}
\,-\,
\frac{1}{2}\, \langle \frac{\delta}{\delta
  \vp} L^{\al_0,\al},\,
\dot {C}^{\al}\star
\frac{\delta}{\delta \vp} L^{\al_0,\al}\rangle  +\,
\partial_\al \log Z^{\al_0,\al} \ .
\label{fe}
\een
Here we use the following notation:
 We write $\,\dot {C}^{\al}\,$ for the derivative 
$\partial_{\al} {C}^{\al_0,\al}\,$, which, as we note,
does not depend on $\al_0$. 
Further, by $\langle\ ,\  \rangle$ we denote the standard scalar product in
$L^2(\mathbb{R}^4, d^4 x)\,$, and $\star$ stands for
convolution in $\mathbb{R}^4$. As an example, 
\ben
\langle\frac{\delta}{\delta \vp},\,\dot {C}^{\al}\star
\frac{\delta}{\delta \vp}\rangle = \int d^4x\, d^4y \ \dot {C}^{\al}(x-y;m)
\frac{\delta}{\delta \varphi(x)} \frac{\delta}{\delta \varphi(y)}
\een
is sometimes called the ``functional Laplace operator''. 
We can now write the flow equation \eqref{fe} in an expanded version as
\ben
\label{CAGFEsexpand}
\begin{split}
&\partial_{\al}{{\cal L}}^{\al_0,\al}_{n}(p_{1},\ldots,p_{n})
= \left({n+2}\atop{2}\right) \, 
\int_{k} \dot{C}^{\al}(k;m)\,{\cal L}^{\al_0,\al}_{n+2}( k, -k,  p_{1},\ldots,p_{n})\\
&-\frac{1}{2}\sum_{n_{1}+n_{2}=n+2} n_{1}n_{2}\   
\mathbb{S}\, \left[  {\cal L}^{\al_0,\al}_{n_{1},l_{1}}(  p_{1},\ldots,p_{n_{1}-1}, q)\,
  \dot{C}^{\al}(q;m)\,   {\cal L}^{\al_0,\al}_{n_{2},l_{2}}(-q,  p_{n_{1}},\ldots,p_{n}) 
\right]\, ,
\end{split}
\een
with $\,q=p_{n_{1}}+\ldots+p_{n}=-p_{1} -\ldots -p_{n_{1}-1}\,$, 
and where $\mathbb{S}$ is the symmetrisation operator acting on 
functions of the momenta $(p_{1},\ldots, p_{n})$ by taking the mean 
value over all permutations $\pi$ of $1,\ldots, n$ satisfying $\pi(1)<\pi(2)
<\ldots<\pi(n_{1}-1)$ and $\pi(n_{1})<\ldots<\pi(n)$. 
We also note that for the theory proposed through \eqref{bare}, 
only even moments (i.e. even in $n,\,n_1,\,n_2\,$) 
will be nonvanishing 
due to the symmetry $\varphi\to-\varphi\,$. Furthermore note
that ${\cal L}^{\al_0,\al}_{2,0}$ vanishes 
identically since the free propagator is absorbed in the covariance,
consistently with \eqref{CAGFEsexpand}. The infinite system of 
equations \eqref{CAGFEsexpand} then constitutes
an  infinite dimensional nonlinear dynamical system.

The CAS functions are defined uniquely as a solution to these 
differential equations 
only after we impose suitable boundary conditions. 
Noting that $L^{\al_{0},\al_{0}}=L_{0}\,$,
these are fixed through the choice of the constants
$c_{0,n}$ in $L_0\,$, \eqref{bare}.
The  CAS functions are then obtained by integrating the
flow equations subject to the boundary conditions.
For an existence and uniqueness
proof in the context of perturbation theory see 
e.g.~\cite{Keller:1990ej,Kop,Muller:2002he}. For 
farther reaching results like e.g. large momentum bounds
~\cite{Kopper:2001to}, bounds on large orders in perturbation
theory~\cite{Kopper:2009um}, applications to finite temperature
field theory~\cite{KMR}, application to nonabelian
gauge theories~\cite{Efremov},  
or a proof of convergence of the operator product 
expansion~\cite{Hollands:2011gf}, we refer
to the respective references. The transition to 
Minkowski space is analysed in ~\cite{KKS}.


\section{The mean field limit of the flow equations}
\label{sec:meanfield}

The flow equations constitute an infinite dimensional
nonlinear dynamical system. The system of functions
$\, {\cal L}^{\al_0,\al}_{n}(p_1,\ldots p_{n})\,$ is 
defined on configuration spaces whose dimension 
also goes to infinity for $\, n\to \infty\,$.  
Since this system is complicated we start analysing 
a simplified dynamical system, where the functions
$\, {\cal L}^{\al_0,\al}_{n}(p_1,\ldots p_{n})\,$ are replaced by
constants  $\, A^{\al_0,\al}_{n}\,$. This amounts to 
setting all external momenta equal to zero in \eqref{CAGFEsexpand}
and to {\it suppose} that the functions
$\, {\cal L}^{\al_0,\al}_{n}(0,\ldots 0, k,-k)\,$
are $\,k$-independent. It thus corresponds to a mean field
limit of the flow equations. There is hope 
that this simplification captures
essential aspects of the behaviour of the full dynamical
system which is in particular based on the fact that the simplification
amounts to replacing the derived propagator in the second 
term on the r.h.s of  \eqref{CAGFEsexpand} by $\,1\,$. In fact we have
\ben
0 \, < \,  \dot{C}^{\al}(q;m)\,=\, e^{-\al(q^2+m^2)}\, \le \, e^{- \al m^2}\, \le \, 1\ .
\label{kleiner1}
\een
So the full system is obtained from the simplified one by 
contracting the second term on the r.h.s. in a momentum dependent
manner. Controlling this contraction seems not to be out of range
though there are hard technical problems, in particular due to the 
fact that the $n$-point functions we want to construct have to respect Bose 
symmetry and euclidean invariance.

A second rather mild simplification will consist in choosing 
$\,m=0\,$ and in restricting in exchange our analysis to the interval
$\al \in [\al_0,\,1]\,$ to avoid infrared problems. It should then be
a straightforward extension of the present analysis
to take the limit $\,\al \to \infty\,$ while keeping $\,m>0\,$. 
Our simplified dynamical system is thus obtained from  
\eqref{CAGFEsexpand} by setting
all external momenta and $m$  equal to zero and setting
\ben
A^{\al_0,\al}_n := {\cal L}^{\al_0,\al}_n(0,\ldots,0)\ .
\label{A}
\een
The system  reads  for $\,n\in 2 \mathbb{N}$
\begin{eqnarray}
\pa_{\al} \, A^{\al_0,\al}_{n} \,=\,
\left({n+2}\atop{2}\right)\, c_\al \
A^{\al_0,\al}_{n+2}
\ -\ \frac12
\sum_{n_1+n_2=n +2}\!\!\!
n_1\,n_2 \ \,A^{\al_0,\al}_{n_1}\,
\,
A^{\al_0,\al}_{n_2}\ ,
\label{dynA}
\end{eqnarray}
where the sum, here and subsequently, is always
over even values of $n_1,\,n_2\,$ only.
Furthermore
\eq
c_{\al}\,:=\,\frac{c}{\al^2}
\ ,\  \mbox{ with } \ \ c \,:=\, \frac{1}{16 \pi^2} 
\label{cal}
\eqe
so that the $\,c_{\al}\,$ is the value at $m\,=\,0\,$ of
\eq
c_{\al}(m)\,:=\,\int_k {\dot C}^{\al}(k;m)\,=\,
\frac{1}{16 \pi^2}\ \frac{1}{\al^2}\ e ^{-\al m^2} \ . 
\label{dcm}
\eqe
As we said \eqref{dynA} is obtained from \eqref{CAGFEsexpand}
by suppressing the momentum dependence of the functions
$\, {\cal L}^{\al_0,\al}_{n}(p_1,\ldots p_{n})\,$ so that
the $\,k$-integral can be carried out explicitly.\\[.2cm]
It is useful to factor out the basic scaling behaviour w.r.t. 
$\al\,$ and combinatoric factors on
setting 
\eq
A^{\al_0,\al}_{n}\,=:\,\al^{n/2-2}\ \frac{1}{n}\  a_{n}(\al)\ ,
\label{Aa}
\eqe
where we suppressed the variable $\,\al_0\,$.
In terms of the functions  $\,a_n(\al)\,$ our dynamical system
can be rewritten as 
\begin{eqnarray}
 a_{n+2}(\al) =
\frac{1}{(n+1)c}\!  \sum_{_{n_1+n_2=n +2}}\!\!\!
a_{n_1}(\al)\, a_{n_2}(\al) \,+\, \frac{n-4}{n(n+1)c} \, a_{n}(\al)
\,+\,\frac{2}{n(n+1)c}\,\al\, \pa_{\al} \,a_n(\al)\ .
\label{dyna}
\end{eqnarray} 
This system permits to construct the functions
$\,a_n(\al)\,$ inductively in $n\,$ if the function $a_2(\al)\,$ is known. 
We make another change of variables  in order to also factor
out the $1/c$ factors
\eq
a_n(\al)\,=\,c^{\frac{2-n}{2}}\  f_n(\al) \ 
\ \mbox{ with the definition }\ \
f_n(\mu) \,:=\, \al^{2-\frac{n}{2}} \ c^{\frac{n-2}{2}}\ n\ 
A^{\al_0,\al}_n\ , 
\label{fA}
\eqe
where we introduced the logarithmic variable $\,\mu \,:=\, 
\ln(\frac{\al}{\al_0})\,$. The system \eqref{dyna} 
can be rewritten
\begin{eqnarray}
 f_{n+2} \,=\,
\frac{1}{n+1} \sum_{n_1+n_2=n +2}
f_{n_1}\, f_{n_2}   
 +\frac{n-4}{n(n+1)} \,  f_{n}
\,+\, \frac{2}{n(n+1)}\, \pa_{\mu} \,f_n \ , \quad \mu \in [\,0,\, 
\ln\frac{1}{\al_0}]\ .
\label{dyna3}
\end{eqnarray} 
Making the functions  $\,f_2\,$ and $\,f_4\,$ more explicit, we can also write
\eq
 f_{4} \,=\,
\frac{1}{3} \,  f_{2}\,(f_2\,-\,1)
\,+\, \frac{1}{3}\, \pa_{\mu} \,f_2 \ ,
\label{f42}
\eqe 
\begin{eqnarray}
 f_{n+2} \,=\,
\frac{1}{n+1} \!\sum_{{n_1+n_2=n +2\atop n_i \ge 4}}
\!\!f_{n_1}\, f_{n_2}   
 +\frac{1}{n+1} \,  f_{n}\,[\,2\,f_2\,+\,1-\frac{4}{n}\,]
\,+\, \frac{2}{n(n+1)}\, \pa_{\mu} \,f_n \ , \quad\! n \ge 4\ .
\label{dyna4}
\end{eqnarray} 
Smooth solutions of the dynamical system
\eqref{f42}, \eqref{dyna4} are fixed if we fix the smooth
function  $\,f_2(\mu)\,$.
In perturbative quantum field 
theory one  primarily considers the flow of the four-point function
which is represented by  $\,f_4(\mu)\,$. 
From \eqref{f42} we realise that we may first fix  $\,f_4(\mu)\,$
and then solve the differential equation \eqref{f42} for 
  $\,f_2(\mu)\,$ to obtain a solution for $\,f_2(\mu)\,$.\\[.2cm]
At this stage we add a few general remarks in relation
with the structure of the system \eqref{f42}, \eqref{dyna4}. 

\begin{itemize}

\item
The first remark concerns
what one might call the {\it combinatorial instability} of the system.  
When trying to figure out an asymptotic behaviour of the $f_n$ as functions
of $\,n\,$, it turns out that, due to the prefactors
the terms 
\[
 f_{n+2}\ ,\quad \frac{1}{n+1}\sum_{{n_1+n_2=n +2\atop n_i \ge 4}} 
\!\!f_{n_1}\, f_{n_2}
\]
are dominant for realistic inductive hypotheses concerning the dependence 
on $\,n\,$, unless one would take into account cancellations of terms of opposite
sign, which typically is beyond scope. As a consequence solving the system by
iteration, starting from a first educated guess and integrating successively,
typically does not define a convergent procedure.

\item
As a consequence of the previous statement we rather proceed in a
different way: We start by fixing $\,f_2(\mu)\,$ and construct the
higher n-point functions from the two-point function. This will
permit to find smooth solutions of the system \eqref{dyna4}, which are interesting 
also from the physical point of view. When asking the question whether  
this procedure is useful for the full system 
\eqref{CAGFEsexpand}, the problem one is faced with is how to
define a function $\,{\cal L}^{\al_0,\al}_{n+2}(p_{1},\ldots,p_{n+2})\,$
 out of the integral 
\[
\int_{k} \dot{C}^{\al}(k;m)\,{\cal L}^{\al_0,\al}_{n+2}( k, -k,  p_{1},\ldots,p_{n})\ ,
\]
once the functions  $\,{\cal L}^{\al_0,\al}_{n'}(p_{1},\ldots,p_{n'})\,$, $n' \le n\,$,
are known. This function has to be 
Bose symmetric, symmetric under the euclidean group, in particular
translation invariant. It also should have  good analyticity 
properties as required  by a full-fledged quantum field theory
which can be analytically continued to Minkowski space. 
A central and presumably hard challenge is to identify the conditions
which determine these functions uniquely in agreement with
 the axioms of quantum field theory.

\item 
We also mention in this context the so-called hierarchy problem
of scalar field theory. It consists in the observation that
in perturbative scalar field theory the two-point function  diverges 
quadratically with the UV cutoff $\,\Lao = \al_0^{-1/2}\,$, as
suggested by \eqref{fA}. In fact it is the only term 
diverging stronger than logarithmically in perturbation theory, 
even when inspecting  the whole of the standard model 
of particle physics. It is then argued that this divergence 
leads to a fine-tuning problem when viewing $\,\Lao\,$ as a 
very high energy scale (``the Planck mass'') since fixing
the mass of the Higgs particle associated to the scalar field 
at its much lower physical value requires fine-tuning of the
corresponding counter term. 
Consequently this quadratic
divergence is often cited as a motivation for supersymmetric 
(or other) extensions of the standard model where 
the perturbative divergences are only logarithmic.
Once we look at the rescaled system \eqref{f42}, \eqref{dyna4} - the same
rescaling can be performed for the full system
\eqref{CAGFEsexpand} - this quadratic divergence disappears.
The  rescaling is natural since it leads to a scale free
system. So from this point of view the hierarchy problem 
appears to be a pseudo-problem of the perturbative treatment, 
whereas on the other hand supersymmetric cancellations appear 
to be due to a subtle fine-tuning procedure.

\end{itemize}


\section{Solutions of the mean field equations}
\label{sec:solutions}

We will consider solutions of \eqref{f42}, \eqref{dyna4} which are smooth functions
of the renormalisation group scale $\,\mu\,$ in the
interval $\,[\,0,\,\ln\frac{1}{\al_0}]\,$.
The existence of the ultraviolet limit means 
that the system of solutions has a finite limit for $\,\al =1\,$
(``when all degrees of freedom have been integrated out'')
when the UV cutoff $1/\al_0\,$ is sent to infinity.
In other words claiming the existence of a mean field solution of the 
FEs in the UV limit is tantamount to prove that  
\ben
\mbox{  the limits } \ \  
\lim_{\mu_{max}\to \infty}\,f_n(\mu_{max}) \ \mbox{ exist for all }\,n \ , 
\quad \mbox{where }\  \mu_{max}\,=\, \ln \frac{1}{\al_0}\  .
\label{limit}
\een

The solutions studied  in \ref{ssec:solundet} are the simplest to obtain.
For these solutions we however do not control the signs of the $n$-point 
functions, not even at $\mu\,=\,0\,$, i.e. for the bare action. We find 
a bare action which is nonpolynomial, and its moments are not 
necessarily positive. 
From the functional integral point of view  the existence of the 
bare action for an arbitrary field  configuration in the support 
of the Gaussian measure is therefore not assured. 
And for the (mean) field configurations for which the bare action exists,
we do not know whether it is uniformly bounded  from below.  
We will show that there exist globally bounded solutions which are monotonically 
increasing as functions of $\,\mu\,$ and vanish  at $\,\mu \,=\, 0\,$ 
when taking the UV limit $\,\al_0 \to 0\,$. They are thus 
asymptotically free in the ultraviolet. The existence of such solutions 
is unexpected from the conventional wisdom point of view.

In section \ref{ssec:asy+} we will then study solutions with strictly
positive boundary conditions at $\,\mu=0\,$ for all $n$-pont functions.
So the bare action is nonpolynomial, and all of its moments have positive
coefficients. The bare action is bounded from below (by $0$) whenever
it is well-defined. The solutions we obtain are again
 ultraviolet asymptotically free. 
Still the bare action (restricted to finite volume) is not well-defined for 
all admissible field configurations since it may diverge due to its nonpolynomial
character.

We therefore study in 
section \ref{ssec:solbounded} solutions the boundary conditions of
which, while being again nonpolynomial, can be resummed into bounded
functions of the field variable and thus lead to well-defined
bare actions in the (finite volume) path integral. These actions are also bounded from
below so that the (regularised)  path integral can be shown to exist.
The solutions from  \ref{ssec:solbounded} constitute
subclasses of those considered in  \ref{ssec:solundet}. 
We show in particular that there exist UV asymptotically free solutions
with well-defined path integral.   
The proof requires much sharper restrictions on the couplings than those
needed in  \ref{ssec:solundet}. 

Finally we study 
the boundary conditions of pure $\vp^4$ theory in section 
\ref{ssec:trivialsol}. The solutions corresponding to these 
boundary conditions  have alternating signs (at least for small
$\mu$) and large $\,\mu$-derivatives which is related to the 
aforementioned fine-tuning of the boundary conditions.     

 We shall find that with the exception of \ref{ssec:solbounded},
the  upper bounds on the coupling 
constants required in the existence proofs of the solutions are 
quite moderate   when compared to constructive field theory upper 
bounds which typically are 
``astronomically small'' (like exponentials of a very big negative
number) due to the high complexity of the contributions
from iterated cluster expansions. The upper bounds from  \ref{ssec:solbounded}
are astronomically  small and not really made explicit. This is because the proof  
of Theorem \ref{meanfieldasyfree} is delicate. So we did not try to optimise 
the bounds w.r.t. the size of the couplings, also for the sake of readability. 
But more reasonable upper bounds should be attainable with reasonable effort.


\subsection{Bounded mean field solutions}
\label{ssec:solundet}

The simplest solutions of \eqref{f42}, \eqref{dyna4} 
are those for which
\eq
 \pa_{\mu} \,f_2\,\equiv\,0\ .
\label{scaleinv}
\eqe
It then follows directly from \eqref{f42}, \eqref{dyna4}
 that
\eq
\, \pa_{\mu} \,f_n\,\equiv\,0 \ \ \forall n 
\eqe
so that we  obtain the $\mu$ independent system
\begin{eqnarray}
f_{4} \,=\,
\frac{1}{3} \,  f_{2}\,(f_2\,-\,1)
\ ,\ \
 f_{n+2} =
\frac{1}{n+1} \!\sum_{{n_1+n_2=n +2\atop n_i \ge 4}}\!\!
f_{n_1}\, f_{n_2}   
 +\frac{1}{n+1} \,  f_{n}\,[\,2 f_2 + 1-\frac{4}{n}\,]
\ , \ \ n \ge 4\ .
\label{dyna5}
\end{eqnarray} 
The solutions of \eqref{dyna5} are fully determined 
on imposing the value of $\, f_{2}\,$.\\[.2cm]

\vspace{-.4cm}

\small 

\noindent
The solutions of \eqref{dyna5} are {\bf scale invariant}, they do not show
any corrections to the canonical scaling factored out in
\eqref{Aa}.
It is  not possible to maintain this condition
beyond the mean field limit since $\,\mu$-independence 
cannot be preserved once we introduce the 
$\,\al$- (and thus $\,\mu$-) dependent propagator $\,\dot C^{\al}\,$.  
So the solutions of \eqref{dyna5} are of limited interest.   
Controlling  them  is essentially trivial. We consider
different cases as regards the value of  $\, f_{2}\,$.\\
\noindent
a) $\ 0 \,<\, |f_{2}|\,\le\, \vep\ll 1\,$\\
In this regime we find that  $\,f_{4}\,=\, {\cal O}(\vep)\,$, with sign opposite
to that of   $\,f_{2}\,$,  $\,f_{6}\,=\, {\cal O}(\vep^2)\,$ with
negative sign, and $\ f_{n}\,>\,0, \ 
f_{n}\,=\, {\cal O}(\vep^2)\,$ for $ n \ge 8\,$. 
So we have an action bounded from below.
This regime is not perturbative, in the sense that
$\,f_{n}\,$ is not of increasing order in $\,\vep\,$ 
for increasing $\,n\,$. The $\,|f_{n}|\,$  for $\,n\,\ge\,6$ 
are bounded by $\,\vep^2\,$ times a numerical 
coefficient becoming small ($\ll 1$) rapidly for increasing $\,n\,$.\\[.1cm]
b)  $\ 0 \,<\, f_{2}\,<\,1\,$\\
In this case  $\,f_{4}\,< 0\,$. Generally
the $\,f_{n}\,$  may have either sign, depending on $\,n\,$. 
\\[.1cm]
c) $\ f_{2}\,=\,1\,$\\
We find   
$\,f_{n}\,=\, 0\,$ $\ \forall$  $\,n\,\ge\,4\,$, i.e.  a ``free theory''.\\[.1cm]
d) $\,f_{2}\,>\, 1\,$\\
By induction on $\,n\,$
one finds that the coefficients  
$\,f_{n}\,$ of the system \eqref{dyna5} satisfy\\
i) $\ f_{n}\,> \,0\ $,   so the action is bounded from below.\\
ii) $\ f_{n}\,$ are strictly increasing when viewed
as functions of  $\,f_{2}\,$ and geometrically bounded by a constant
to the power $\,n\,$.\\
In particular for $\,f_{2}\,=\, 1+\vep\,$ with $0 < \vep \ll 1\,$ we find  
$\,f_{4}\,=\, \frac13\,(1+\vep)\vep \,$,
$\,f_{6}\,=\, \frac{1}{15}\,(1+\vep)^2\vep \,$,
$\ f_{n}\,=\,O(\vep)\,$  $\ \forall$  $\,n\,\ge\,4\,$.
The $\,|f_{n}|\,$ for $\,n\,\ge\,6$  are bounded
by $\,\vep^2\,$ times a numerical 
coefficient becoming small ($\ll 1$) rapidly for increasing 
$\,n\,$.\\[.1cm]  
e) $\,f_{2}\,<\, 0\,$\\
In this case we do not control the signs of the $\,f_n\,$.
The   $\,|f_{n}|\,$ may become large in modulus for large $\,f_{2}\,$.

\normalsize

\vspace{.3cm}

We now study more general solutions for which
all $|\,f_n\,|$ are bounded by 1. 
We consider a smooth two-point function satisfying
for $\,0 \,< \,\de\,<\,1\,$
\eq
-K_1 \,\de\,\le \, f_2(0)\,\le\, -\de\,  < \, 0 \, ,\ 
\ |\,\pa^{l}_\mu \,f_2(\mu)\,|\, \le\, \frac{K_1^{l}\,\de^{l+1}}{(l+1)^2}\,
 l\, ! \quad \forall\ \mu \in [\,0,\, \ln \frac{1}{\al_0}\, ]
\ \mbox{ and }\ \forall\  l \ge 0\  ,
\label{f2d}
\eqe
where $\,K_1\,>1 \,$ is a  positive constant. 
We restrict ourselves
for simplicity and definiteness to the interval $\,0 < \de \,<\, 1\,$,
but larger values could  be analysed similarly.
We note that the sign of $\,f_2(0)\,$ in \eqref{f2d}
is in agreement with the 
sign of the mass counter term in perturbative $\vp^4$-theory
at lowest order. 

{\prop 
\label{geom2}
For suitable $\,K \,\ge\,\sup(K_1,\,4)\,$ 
and  $\,f_2(\mu)\,$
satisfying \eqref{f2d}, the functions  
$\,f_n(\mu)\,$ solving \eqref{dyna4} are smooth and satisfy
for $\,\mu \in [0, \ln\frac{1}{\al_0} ]\,$,\
$\,n \in 2\mathbb{N}+2\, , \ \, l \ge 0\,$ 
\eq
|\pa_\mu^{\,l}\, f_n(\mu)| \ \le \
\frac{K^{n+l-2}\,\de^{l+1}\,}{(l+1)^2}\ \frac{(n+l-2)!}{(n-2)!} \ .
\label{undetbound}
\eqe
}
\vspace{.1cm}

\noindent
{\it Proof}.
The proof is by induction in $n+l \ge 2\,$.\\
The bounds hold for the two-point function by assumption
\eqref{f2d}. 
Verification of the bounds on $\,\pa_\mu^{\,l}\, f_4(\mu)\,$
using  \eqref{f42} is straightforward and simpler than
the general case $\,n \ge 4\,$. So we leave this case to the reader. 
For $\,n \ge 4\,$ we insert the induction hypothesis on the r.h.s.
of \eqref{dyna4},
derived $l\,$ times w.r.t. $\mu\,$. This gives the bound 
\ben
\begin{split}
 & {\quad\ }
\frac{\de^{l+2}}{n+1} \sum_{{n_1+n_2=n +2, n_i \ge 4\atop l_1+l_2=l}}
{l \choose l_1}\,  
\frac{K^{n+l-2}}{(l_1+1)^2 \ (l_2+1)^2 }\ 
\frac{(n_1+l_1-2)!}{(n_1-2)!}\  
 \frac{(n_2+l_2-2)!}{(n_2-2)!}\\
& +\,\frac{\de^{l+1}}{n+1} \,\sum_{l_1+l_2=l} {l \choose l_1}\,  
\frac{K^{n+l_1-2}}{(l_1+1)^2}\ \frac{(n+l_1-2)!}{(n-2)!}  
\,[\,\de\,\frac{K^{2+l_2-2}}{(l_2+1)^2}\, 2\, (2+l_2-2)! 
+ \de_{l_2,0}\,(1-\frac{4}{n})\ ]\\
&\,+\, \frac{2\,\de^{l+2}}{n(n+1)}\, 
\frac{K^{n+l+1-2}}{(l+2)^2}\ \frac{(n+l+1-2)!}{(n-2)!} 
\ .
\label{ddynabd3}
\end{split}
\een
Using the standard bound (all entries are supposed to be nonnegative
integers) 
\eq
{l \choose l_1}\, {n-2 \choose n_1-2} \le {n-2+l \choose n_1-2 + l_1} 
\label{standard1}
\eqe
we obtain the following estimate for \eqref{ddynabd3}
\ben
\begin{split}
 & {\quad\ }
\frac{\de^{l+2}}{n+1} \sum_{{n_1+n_2=n +2, n_i \ge 4\atop l_1+l_2=l}}
\frac{K^{n+l-2}}{(l_1+1)^2 \ (l_2+1)^2 }\ 
\frac{(n+l-2)!}{(n-2)!}  
\\
& +\,\frac{\de^{l+1}}{n+1} \,\sum_{l_1+l_2=l}   
\frac{K^{n+l-2}}{(l_1+1)^2}\ \frac{(n+l-2)!}{(n-2)!}  
\,[\,\frac{2\,\de\,}{(l_2+1)^2}\
+ \de_{l_2,0}\,(1-\frac{4}{n})\ ]
\label{dynbd5}
\\
&\,+\, \frac{2\,\de^{l+2}}{n(n+1)}\, 
\frac{K^{n+l+1-2}}{(l+2)^2}\ \frac{(n+l+1-2)!}{(n-2)!} 
\ .
\end{split}
\een
Choosing $K$ sufficiently large  such that for $\,n \ge 6\,$ 
\[
\frac{\de}{n+1} \! \!  \sum_{{n_1+n_2=n +2, n_i \ge 4 \atop l_1+l_2=l}}
\frac{1}{(l_1+1)^2\, (l_2+1)^2 }\ \le\ \frac{1}{3}\
\frac{K^2}{(l+1)^2}\  \frac{(n+l)(n+l-1)}{n(n-1)} 
\]
and such that
\[
\frac{1}{n+1} \! \!  \sum_{l_1+l_2=l} 
\frac{1}{(l_1+1)^2\, (l_2+1)^2 }\ 
\,[\,2 \de
+ \de_{l_2,0}\,(1-\frac{4}{n})\ ]
\ \le \ \frac{K^2}{2\, (l+1)^2}\  \frac{(n+l)(n+l-1)}{n(n-1)} 
\]
and such that 
\[
\frac{2\, \de}{n(n+1)}\, \frac{1}{(l+2)^2}\  \le \  \frac{1}{6}\,
K \ \frac{1}{(l+1)^2}\
\frac{n+l}{n(n-1)} 
\]
we find that \eqref{ddynabd3}  is bounded by
\[
( \frac{1}{3}\,+\, \frac{1}{2}\,+\, \frac{1}{6})\,
\frac{\de^{l+1}\,K^{n+l}}{(l+1)^2}\ \frac{(n+l)!}{n!}\ \, .
\]
One can straightforwardly 
convince oneself $\,K\,=4\,$ is admissible for 
$\,\de\,=\,1\,$, $\,K_1 \le\,4\,$ and that smaller
values of  $\,K\,$ are allowed if $\,\de\,< \,1\,$, 
 $\,K_1\, <\,4\,$. 
\qed

\vspace{.2cm}

Going back to the dynamical system \eqref{dynA} we obtain from the
set of smooth  functions $\,f_n(\mu)\,$ the system of   smooth
 functions  $\,A^{\al_0,\al}_{n}\,$.
If the functions  
  $\,f_n(\mu)\,$  satisfy the bounds from Proposition \ref{geom2}, then
the  $\,A^{\al_0,\al}_{n}\,$
satisfy
the bounds
\eq
| A^{\al_0,\al}_{n}| \,\le\, \de\ 
\left(\frac{ \al\, K^2}{c}\right)^{\frac{n-2}{2}}\ 
\frac{1}{\al n}  \quad
\mbox{ for }\ 0  < \al_0 \le \al \le 1\ .  
\label{dAbound}
\eqe
This bound is uniform in $\al_0$.

\vspace{.2cm}

\noindent
In order to show that there is a subclass of solutions among 
those from Proposition \ref{geom2} 
which describe a mean field theory in the sense that the solutions 
$\,f_n(\mu_{max})\,$ have a well-defined limit for $\,\al_0 \,\to\, 0\,,\ 
\mu_{max}\,\to\, \infty\,$,
we choose
\ben
\! f_2(\mu)\,=\,-\de(\mu)\, ,\ \,  \de(\mu_{\max})=\,\de\, , \ \, 
\pa_\mu\,  \de(\mu)=\, \be\, \de^2(\mu)\, , \ \, 0 \,<\,\de\,,\; \be \,<\,\frac12\, ,
\  \, \mu \in [0,\,\ln\frac{1}{\al_0}\,]\ .
\label{f2asy1}
\een 
The well-known solution is
\ben
\de(\mu)\,=\, \frac{\de}{1\,+\,  (\mu_{\max}-{\mu})\be\,\de\,}\ \  .  
\label{f2asy2}
\een
Evidently \eqref{f2asy1} verifies 
the assumptions of Proposition \ref{geom2}. We have in 
particular 
\ben
\lim_{\mu_{max}\to \infty} f_2(\mu_{max})\,=\, -\,\de\ ,\quad
\lim_{\mu_{max}\to \infty} \pa_\mu^l\,f_2(\mu_{max})\,=\, -\,\be^l\, l!\ \de^{l+1}\ \,.
\label{limits}
\een 
By straightforward induction in $\,n\,+\,l\,$, proceeding as 
in the proof Proposition \ref{geom2} we then find that the limits
$\,\lim_{\mu_{max}\to \infty} f_n(\mu_{max})\,$, $\,n\ge 4\,$, also exist and obey 
the bounds of Proposition \ref{geom2}. We collect our findings in 
{\prop 
\label{geom3}
Among the solutions from Proposition \ref{geom2} there are 
nontrivial asymptotically free solutions, for which hold
\ben
f_2(\mu_{max})\, < 0\ ,\quad f_4 (\mu_{max})\, >\,0 \ \ ,
\label{1}
\een
\ben
\lim_{\mu_{max}\to \infty} f_n(\mu_{max})\quad \mbox{exists}\ 
\quad \forall \,n\, \in 2\mathbb{N} \ \ ,
\label{2}
\een
\ben
\lim_{\mu_{max}\to \infty} \pa_\mu^l\,f_n(0)\,=\, 0 
\quad \forall \,n\, \in 2\mathbb{N}\,,\ \, l \in \mathbb{N}_0 \ \ .
\label{3}
\een
}

\noindent
{\it Proof}. The second inequality in \eqref{1} is true
if $\,\frac13\ \de(\mu_{max})(\de(\mu_{max})+1) - \pa_\mu \de(\mu_{max}) > 0\,$,
which is the case for $\,\de,\, \be\,$ bounded as in \eqref{f2asy1}. The last 
statement \eqref{3} again follows by induction proceeding as in the proof
of Proposition \ref{geom2}. 
\qed

\vspace{.2cm}

We also note  that solutions of the type \eqref{f2asy1}, but with 
  $\,\be\,<\,0\,$ negative, will lead to trivial theories, namely 
we find that
\ben
\lim_{\mu_{max}\to \infty} f_n(\mu_{max}) \,=\,0
\label{trivun}
\een
if $\,\de(0)\,$ is fixed to be positive and not too large. 
In fact one obtains in this case 
\ben
\de(\mu)\,=\, \frac{\de(0)}{1\,-\,{\mu}\,\be\,\de(0)\,}  
\label{f2triv2}
\een
which vanishes for $\mu\,=\,\mu_{max}\to \infty\,$ and 
then implies the vanishing of all $\,f_n\,$ in this limit,
which can be proven  again inductively as 
Proposition \ref{geom2}.  We do not work out this point further here.
We will come back to the triviality question in \ref{ssec:trivialsol}.

\subsection{Mean field solutions with positive bare values}
\label{ssec:asy+}

The bare actions constructed from  
the solutions $\,f_n(0)\,$ in \ref{ssec:solundet}
are generally not bounded from below.
In this subsection we look at solutions for which 
all $\,f_n\,$ are positive and monotonic
\eq
f_n(\mu) \ge 0\ ,\quad \pa_{\mu} f_n(\mu) \ge 0\ .
\label{reqprop}
\eqe
These properties assure positivity
of bare action whenever it is well-defined. 

\vspace{.1cm}

\noindent
We obtain smooth solutions of \eqref{f42}, \eqref{dyna4} satisfying the conditions
\eqref{reqprop} on  considering functions $\,f_2(\mu)\,$ such that 
\eq
 f_{2}(\mu) \,=\, 1 +\de(\mu)\ ,\ \  0\,<\,\de(\mu)\,< 1\ ,\ \
 \pa_{\mu} \,\de \,=\, \beta\, \de^2(\mu) \ ,
\ \ 0 \,<\, \be \,<\,1\ , 
\ \ \mu \in [\,0,\, \ln \frac{1}{\al_0}\, ]\ .
\label{f2}
\eqe
The main difference between \eqref{f2} and \eqref{f2asy1}
 is that $\,f_2\,$ in \eqref{f2} is not of 
order $\,\delta\,$. So the solutions
studied are nonperturbative from the beginning.

\noindent
Rewriting \eqref{f42} as
\eq
3\,f_4(\mu)\,=\, (1 +\de(\mu))\, \de(\mu) \,+\, \pa_\mu \de(\mu)\ ,
\label{f4}
\eqe
we see that the relations \eqref{f2}, \eqref{f4} imply
\eq
\pa^{\,l}_{\mu} \, f_4(\mu)\,\ge \,0 \quad \forall l \ .
\label{f4n}
\eqe

{\prop 
\label{geom}
For suitable $\,K > 1\,$ and $\,f_2(\mu)\,=\,1 \,+\,\de(\mu)\,$ smooth, 
satisfying \eqref{f2}, the functions  
$\,f_n(\mu)\,$ are smooth, and satisfy
for $\,\mu \in [\,0, \ln\frac{1}{\al_0} ]\,$,\
$\,n \in 2\mathbb{N}+2\, ,\ \, l \ge 0\,$ 
\eq
0 \ <\ \pa_\mu^{\,l}\, f_n(\mu) \ \le \
\de(\mu) \ \frac{K^{n+l-2}}{(l+1)^2}\ \frac{(n+l-2)!}{(n-2)!} \ .
\eqe
}
\vspace{.1cm}

\noindent
{\it Proof}.
The proof is by induction, in $n+l \ge 2\,$.
Positivity follows immediately by inspecting the
r.h.s of \eqref{dyna4}. The bound does not contain 
higher powers of $\delta(\mu)\,$ as in Proposition \ref{geom2}
since $f_2$ is no more 
of order $\delta$.
Otherwise the proof follows strictly that of Proposition
\ref{geom2}. So we do not rewrite it. We find again
that $\,K=4\,$ is an admissible value.
\qed

\vspace{.2cm}

From \eqref{dynA} and Proposition \ref{geom} we find
bounds for 
the $\,A^{\al_0,\al}_{n}\,$ 
\eq
0 \ < \  A^{\al_0,\al}_n \ < \ ( \de_{n,2}+ \de(\mu))\
\left(\frac{ \al\, K^2}{c}\right)^{\frac{n-2}{2}}\, 
\frac{1}{\al \,n}  \quad
\mbox{ for }\ 0  < \al_0 \le \al \le 1  
\label{Abound}
\eqe
which are uniform in the UV cutoff $\,\al_0\,$.
The initial data  $\,f_n(0)\,\ge \,0\,$
assure the positivity of all moments of the bare action,  
which obey the bounds \eqref{Abound}.\\
The solutions studied in Propositon \ref{geom} are again
asymptotically free. When choosing $\,\de\,=\,\de(\mu_{max})\,>\,0$ 
fixed, we have  statements analogous to  
(\ref{f2asy2}, \ref{limits}, \ref{1}, \ref{2}, \ref{3})\,:
\ben
\de(\mu)\,=\, \frac{\de}{1+ (\mu_{\max}-{\mu})\be\,\de\,}\ \  .  
\een
\ben
\lim_{_{\mu_{max}\to \infty}} \pa_\mu^l\,f_2(\mu_{max})\,=\, 
\de_{l,0}\,+\,\be^l\, l!\, \de^{l+1}\, ,
\ \ 
\lim_{_{\mu_{max}\to \infty}} f_4(\mu_{max})\,=\, \frac13
(1 +\de)\de\,+\, \be \de^2\ ,
\label{11}
\een
\ben
\lim_{\mu_{max}\to \infty} f_n(\mu_{max})\quad 
\mbox{exists and is positive}\ 
\quad \forall \,n\, \in 2\mathbb{N}\  \ ,
\label{22}
\een
\ben
\lim_{\mu_{max}\to \infty} \pa_\mu^l\,f_2(0)\,=\, \de_{l,0} \ ,\quad
\lim_{\mu_{max}\to \infty} \pa_\mu^l\,f_n(0)\,=\, 0 
\quad \forall \,n\, \in 2\mathbb{N}+2 ,\ l \in \mathbb{N}_0 \ \ .
\label{33}
\een

\vspace{.3cm}

The asymptotically free solutions we have considered
 seem to be 
quite special. Still, as regards the UV limit, the basic
possibilities are nontrivial asymptotically free
or safe (i.e. scale independent) solutions, or trivial 
solutions which are free at $\,\mu\,=\,\mu_{max}\,$. 
An interesting task left for the future is to analyse the 
different classes of solutions more systematically.    

\vspace{.2cm}

\small

\noindent
In perturbative quantum field theory one generally analyses
in a first place the scaling behaviour of the four-point
function, and not that of the two-point function. 
When imposing $\,f_4(\mu)\,$ and then solving the differential 
equation for the two-point function \eqref{f4}, which is of Riccati
type, we find as a particular solution 
\eq
f_2 (\mu)= - 3 f_4(\mu)\ .
\label{199}
\eqe
This implies that we can find all solutions of the Riccati
equation. For example, the one satisfying $f_2(0) \,=\, 0\,$ is given
by
\eq
f_2(\mu)\,=\,\frac{3\, f_4(0)\,e^{\int_0^\mu(6 f_4(\mu')+1)\,d\mu'}}{1
\,+\, 3\, f_4(0)\,
\int_0^\mu d\mu'\  e^{\int_0^{\mu'}(6 f_4(\mu'')+1)\,d\mu''}}\ - \,  
3\, f_4(\mu) \ .
\label{f200}
\eqe
It satisfies 
\[
f_2(\mu) \ge 0\ ,\quad  \pa_{\mu} f_2(\mu)\,\ge \,0\ ,\quad
f_2(0)\,=\,0\  .
\]   
Higher order derivatives of this solution are however not positive
for all values of $\,\mu\,$. It is therefore not clear whether the 
corresponding solutions 
of \eqref{dyna4} satisfy Proposition \ref{geom} for $\,l=0\,$. 
Since the solution \eqref{f200}
has  a vanishing mass counter term, 
it may well be that the ultraviolet limit for this solution
does not exist.
 
An interesting project would be to find out whether one
can  construct along these lines asymptotically
free solutions, in particular for the four-point function, 
satisfying Proposition \ref{geom} which are such that the bare action
is well defined and bounded from below.

\normalsize


\subsection{Solutions of bounded action} 
\label{ssec:solbounded}

The bare actions of the solutions constructed 
so far  are nonpolynomial and generally not well-defined 
on the whole of the support of the Gaussian measure.
The solutions we will construct in this 
section satisfy
sufficiently strong bounds in order to assure well-defined 
bare actions bounded from below. Since the estimates become 
more delicate the upper bounds on the coupling constants 
required are much more restrictive.
The solutions from this section will in fact be 
subclasses of those from section \ref{ssec:solundet}.

\vspace{.2cm}

\small
\noindent
The bare functional \eqref{bare} 
has the form 
$\ L_0(\vp)\,=\, \sum_{n\in 2\mathbb{N}} \int d^4x \ c_{0,n}\, \vp^n(x)\ .$
The constants $\,c_{0,n}\,$ are related to the $\,A_n(\al_0)$ via 
\[
 c_{0,n}\,=\, \frac{1}{(2\pi)^{4}}\, A_n(\al_0)
\]
as can be seen from \eqref{c0}, \eqref{deltaCAG}, \eqref{A}.
For the $\,A_n(\al_0)\,$ we have deduced the bounds
\eqref{dAbound} resp. \eqref{Abound}.
The functional $\,L_0(\vp)\,$ is well-defined  for all $\,\vp\,$ in  
\[
{\cal D}(\al_0)\,=\, \left\{ \vp \ \Big| \ \vp 
\in  \bigcap_{n\in2\mathbb{N}}\ L^{n}(\mathbb{R}^4,d^4x)\ , \ \ 
L_0(\vp)\,<\, \infty\,\right\} \  .
\]
If the bounds \eqref{dAbound} resp. \eqref{Abound} hold,
the set $\,{\cal D}(\al_0)\,$
contains
\[ 
{\cal M}_{\vep}(\al_0)\,:=\,\left\{ \vp\ \Bigg| \ \vp \in  \bigcap_{n\in2\mathbb{N}}  
L^{n}(\mathbb{R}^4,d^4x)\ , \quad
\lim\sup ||\vp||_{n} \,< \,(\frac{c}{\al_0\,K^2})^{1/2}\,-\vep  
\right\}\ \subset {\cal D}(\al_0)
\]
for arbitrarily small positive $\,\vep\,$.
The sets $\,{\cal M}_{\vep}(\al_0)\,$  do not exhaust the support of the measure
$\,\mu(C^{\al_0,\al})\,$ for finite $\,\al_0\,$. 
One might then be tented to introduce one more regularisation
by setting 
\[
 \quad \qquad V(\vp)\,\equiv\,  e^{-L_0(\vp)}\ \ ,\quad \mbox{if} \ \ \vp \in  
\mbox{supp}\,\mu(C^{\al_0,\al})\cap\ {\cal D}(\al_0)\ ,
\]
\[
\quad\qquad V(\vp)\,\equiv\,0  \quad\qquad ,\ \  \, \mbox{ if} \ \ \vp 
 \in \mbox{supp}\,\mu(C^{\al_0,\al})-{\cal D}(\al_0) \ \, .\,\,
\]
But $\,V(\vp)\,$ is not differentiable
w.r.t. $\,\vp\,$, and it is thus no more possible to  derive the FEs
 from the path integral by partial integration. 
In fact boundary terms appear where the potential 
 $\,V(\vp)\,$ is cut off.

\normalsize

\vspace{.2cm}

\noindent
To impose boundedness from the beginning
we instead study bare actions, still supposed to be local, which are 
of the form
\eq
\boxed{\ L_0(\vp)\,=\, \sum_{n\in 2\mathbb{N}} \int d^4x\  
{ \cal \ti L}_n \ \sin(\al_0^{n/2}\vp^n)\ \al_0^{-n/2}\ }\ . 
\label{L0sin}
\eqe
Since we are interested in the bare action we only consider
the FEs for the functions $\,{\cal \ti L}_n\,$ and their 
$\al$-derivatives evaluated at $\,\al=\al_0\,$. 
When going to zero external momentum they take the form 
 \ben
\label{fesin}
\begin{split}
&
\  \sum_{{\nu \ge 0 \atop (1+2\nu) n' =n}} \al_0^{n'\nu}\,
\frac{(-1)^{\nu}}{(2\nu+1)!}\, \pa_{\al} { \cal \ti L}_{n'}\ +
\sum_{{\nu \ge 1 \atop (1+2\nu)n' =n}} 
\al_0^{\nu n'-1}\,\nu n' \, \frac{(-1)^{\nu}}{(2\nu+1)!} \,
{ \cal \ti L}_{n'}\\ 
&\,=\,
\frac{c}{2} (n+2)(n+1)\,\al_0^{-2}\, { \cal \ti L}_{n+2}\,+\,
\frac{c}{2}\,\al_0^{-2}\,\sum_{{\nu \ge 1 \atop (1+2\nu) n' =n+2}}
 \al_0^{\nu n'} \,  n'(n'-1)\,
\frac{(-1)^{\nu}}{(2\nu)!} \,{ \cal \ti L}_{n'}\\
&\,+\,
\frac{c}{2}\,\al_0^{-2}\,\sum_{{\nu \ge 1 \atop (1+2\nu) n' =n+2}}
\al_0^{\nu n'} \,   n'^{\,2}\,
\frac{(-1)^{\nu}}{(2\nu-1)!}\, { \cal \ti L}_{n'}\\
&\,-\,\frac12 \sum_{_{\footnotesize{\substack{n_1+n_2 = n+2,\ \nu_1,\nu_2\ge 0\\
(1+2\nu_1) n'=n_1\\ (1+2\nu_2) n''=n_2}}}}   
\al_0^{n' \nu_1\,+\,n'' \nu_2}\
n_1\,n_2\ 
\frac{(-1)^{\nu_1+\nu_2}}{(2\nu_1)!(2\nu_2)!}\, 
{ \cal \ti L}_{n'}\, { \cal \ti L}_{n''}\,
\ .
\end{split}
\een
The additional terms appearing as compared to \eqref{CAGFEsexpand} 
and \eqref{dynA},  carry $\nu \ge 1\,$ in the various sums. 
They stem from higher order terms on expanding the sine function
in \eqref{L0sin}.
So \eqref{fesin} follows directly from 
\eqref{CAGFEsexpand}, \eqref{dynA}, when expressing
the ${\cal L}_n$ in terms of the ${\cal \ti L}_{n'}\,$.
We note that
\ben
{\cal \ti L}_{2}\,=\,{\cal L}^{\al_0,\al_0}_{2}\ .
\label{ti2}
\een
As before we introduce dimensionless functions $\,a(n)\,$
via the definition
\ben
 { \cal \ti L}_n\,=:\,\frac{1}{n} \  \al_0^{n/2-2}\ a(n)  \ .
\een
This gives the  following FEs for the $\,a(n)\,$, 
evaluated at $\al=\al_0$ or equivalently at $\,\mu=0\,$
 \ben
\label{ffesin}
\begin{split}
 a(n+2) 
&\,=\, \frac{1}{n+1}\, \sum_{{\nu \ge 1 \atop (1+ 2 \nu)n' =n+2}} 
(n'-1) \frac{(-1)^{\nu-1}}{(2\nu)!}\, a(n')
\,+\,
\frac{1}{n+1}\, \sum_{{\nu \ge 1 \atop (1+ 2 \nu)n' =n+2}} 
n'\, \frac{(-1)^{\nu-1}}{(2\nu-1)!}\, a(n')\\
&\,+\,\frac{1}{(n+1)c} 
\sum_{_{\footnotesize{\substack{
n_1+n_2 = n+2,\ \nu_1,\nu_2\ge 0\\
(1+2 \nu_1) n'=n_1\\ (1+2 \nu_2) n''+n_2} }}}  
(-1)^{\nu_1+\nu_2}\frac{(1+2\nu_1)(1+2\nu_2)}{(2\nu_1)!(2\nu_2)!}\,
 a(n')\, a(n'')\\
&\,+\,\frac{2}{(n+1)c}\,\frac{n-4}{2n}\,a(n)\,+\,
\frac{2}{(n+1)c}\, 
\sum_{{\nu \ge 1 \atop (1+2\nu)n' =n}} 
\frac{n'-4}{2n'}\,\frac{(-1)^{\nu}}{(2\nu+1)!}\,a(n')\\
&\,+\,
\frac{2}{(n+1)c}\,\frac{1}{n}\,\pa_\mu a(n)\,+\,
\frac{2}{(n+1)c}\,\sum_{{\nu \ge 1 \atop (1+2\nu)n' =n}} \frac{1}{n'}\,
\frac{(-1)^{\nu}}{(2\nu+1)!}\,\pa_\mu a(n')\\
&
\,+\,
\frac{2}{(n+1)c}\sum_{{\nu \ge 1 \atop (1+2\nu)n'=n}}
 (-1)^{\nu}\ \frac{\nu}{(2\nu+1)!}\,\ a(n')
\ .
\end{split}
\een
For $n=2$ we obtain simply
\ben
 a(4)\,=\, \frac{1}{3\,c}\left[a(2)(a(2)-1) \,+\, \pa_\mu a(2)\right]
\label{a4l}
\een
in agreement with \eqref{dyna} for $n=2\,$.\\
We write for shortness 
\eq
a(n,l):= \pa_\mu^l\, a(n)\ .
\eqe
Our bound will be expressed in terms of the decomposition
of $\,n\,$ in prime numbers.
We write this decomposition for general $\,n\,$ in hopefully obvious notation as
\eq
n\,=\, 2^{p_2(n)}\,\cdot\,  3^{p_3(n)}\,\cdot\, 5^{p_5(n)}\,\cdot\,
7^{p_7(n)}\,\cdot\,  11^{p_{11}(n)}\ \ldots
\eqe
We also define 
\ben
\label{Bs}
\begin{split}
&B(n,0)\,:=\,B(n)\,:=\,  \ [\,2^{p_2(n)/4}\,\cdot\, 
 3^{p_3(n)/4}\,\cdot\, 5^{p_5(n)/2}\,\cdot\,7^{9\,p_7(n)/8}
\,\cdot\,11^{9p_{11}(n)/8}\,\ldots\ ]^{-1} \ \; ,\\
&
B(n,l)\,:=\, B(n)\ \frac{(n+l)!}{n!}\ \; ,  \quad 
B_\vep(n,l)\,:=\, B(n,l)\,\vep^{l+1} \ \; .
\end{split}
\een

\vspace{.2cm}

{\lemma 
For fixed   $\,\vep >0\,$ sufficiently small, 
there exists  $\,\vep'$  satisfying $\,0 \,<\,\vep' \,\le\, \vep\,$  
such that if   
\eq
\left|\,a(2,l)\,\right| \,\le \ B_{\vep'} (2,l)  
\quad \forall\ l\,\ge\, 0
\ , 
\label{a2l}
\eqe
then   
\eq
\left|\,a(n,l)\,\right| 
\,\le \  B_\vep(n,l) \quad \forall\ l\,\ge\, 0\ , \ n \,\ge\, 2\ .
\label{anl}
\eqe
\label{boundedaction}
}

\noindent
{\it Proof}.
We proceed by induction in $\,N\,=\,n\,+\,l\,\ge \,2\,$, going up
in $\,l\,$ for fixed $\,N\,$. We are not 
ambitious on the size of $\,\vep\,$, $\,\vep'\,$. 
On inspecting \eqref{a4l}, \eqref{ffesin} it is obvious 
that the bound \eqref{anl} holds for all $(n,\,l)$ with 
$n\,+\, l \,\le\, N_0\,$, for  $\,N_0$ fixed and $\,\vep'\,$
chosen sufficiently small depending on $\,N_0\,$ and $\,\vep\,$.     
We will not derive an explicit upper bound on
$\,\vep'(\vep,N_0)\,>\,0\,$, satisfying 
ourselves with the existence statement,  but
 comment on the size of $\,N_0\,$ in the proof. 
To improve the upper bounds on $\,\vep\,$, $\,\vep'\,$
one has to consider small values of $\,N\,$ explicitly,
and  to bound separately 
particular cases, where $\,B(n+2)\,$ is much bigger
than $\,B(n)\,$. This  is the case for example for
$\,n+2 =2^k\,$ with $\,k\,$ large, where
$\,B(n+2)\,=\, (n+2)^{-1/4}\,$ whereas  $\,B(n)\,$ may be 
smaller than $\,1/n\,$.\\[-.3cm]
     
\noindent
We consider \eqref{ffesin} and bound inductively the
l.h.s. of this equation in terms of the r.h.s.
We will treat  $\mu$-derivatives of \eqref{ffesin} afterwards.  
We bound successively the terms on the r.h.s. of \eqref{ffesin}.

\begin{itemize}

\item{1st term}
\ben
\label{1stterm}
\begin{split}
&\Biggl|\, \frac{1}{n+1}\! \sum_{{\nu \ge 1 \atop (1+ 2 \nu)n' =n+2}} 
(n'-1) \frac{(-1)^{\nu-1}}{(2\nu)!}\, a(n') \,\Biggr|\
\le\
\frac{n+2}{n+1}\!\! \sum_{{\nu \ge 1 \atop (1+ 2 \nu)n' =n+2}} \!\!
\frac{1}{1+2\nu}\
\frac{\left|\,a(n')\,\right|}{(2\nu)!}\\
 \le &\
\frac{n+2}{n+1}\, \left(\frac{3^{1/4}}{3!}+\frac{5^{1/2}}{5!}+
\frac{7^{9/8}}{7!}+\frac{9^{1/4}}{9!}+\ldots\right)B_\vep(n+2)\\ 
 \le &\ \frac{n+2}{n+1}\, 
\left(0.22\,+\,0.02\,+\,0.002\,+\,\ldots \right) \, B_\vep(n+2)\,\,
\le \ 0.25\ \frac{n+2}{n+1}\   B_\vep(n+2)\ .
\end{split} 
\een  
In the second line we use the identity 
\[
B_\vep(n')\,=\,   (1+2\nu)^{p_{\nu}}\  
B_\vep(n+2)\ \mbox{ for }\ n+2\,=\,(1+2\nu)\,n'\ .
\]
Here the exponent $\,p_{\nu}\,$ is to be read from the definition
\eqref{Bs}. It varies between $1/4$ and $9/8\,$ depending on the
value of $1+2\nu\,$.

\item{2nd term}

\ben
\label{2ndterm}
\begin{split}
&\Biggl|\, \frac{1}{n+1}\!\! \sum_{{\nu \ge 1 \atop (1+ 2 \nu)n' =n+2}}\!\! 
n'\, \frac{(-1)^{\nu-1}}{(2\nu-1)!}\, a(n')\,\Biggr|\\
 \le\, &\, 
\frac{n+2}{n+1}\, 
 \left( 3^{-3/4}\,+\, \frac{5^{-1/2}}{3!}\,+\, 
\frac{7^{1/8}}{5!}\,+\,\frac {9^{-3/4}}{7!}\,+\,
\ldots\right)\,B_\vep(n+2)\,\\ 
\le\,&
 \left(0.44\,+\,0.075\,+\,0.011\,+\,\ldots \right) \ \frac{n+2}{n+1}\ 
B_\vep(n+2)\,\le\,  
\ 0.53\ \frac{n+2}{n+1}\ B_\vep(n+2)\ .
\end{split} 
\een

\item{3rd term}
\ben
\label{3rdterm}
\begin{split}
&\Biggl|\, \frac{1}{(n+1)c} 
\sum_{{\footnotesize\substack{n_1+n_2 = n+2\\ \nu_1,\nu_2\ge 0\\
(1+2 \nu_1) n'=n_1\\
 (1+2 \nu_2) n''=n_2} } }  
(-1)^{\nu_1+\nu_2}\frac{(1+2\nu_1)(1+2\nu_2)}{(2\nu_1)!(2\nu_2)!}\,
 a(n')\, a(n'')\,\Biggr|\\
&
 \,\le\,
\frac{\vep^2}{(n+1)c}\, 
\sum_{\nu_1,\nu_2\ge 0}\frac{(1+2\nu_1)(1+2\nu_2)}{(2\nu_1)!(2\nu_2)!}\ \
2\sum_{n_1 \le n_2}\
\sum_{n' \le \frac{n+2}{2(1+\nu_1)}}
B(n')\,[\frac{1}{2}\frac{n+2}{1+2 \nu_2}]^{-1/4}\ .
\end{split} 
\een  
We have
\eq
\label{fn1}
\!\sum_{n' \le \frac{n+2}{2(1+\nu_1)}}\!
\! B(n')\,\le \!\!
\sum_{\mu,\nu,\rho\ge 0} 2^{-\frac{\mu}{4}}\,3^{-\frac{\nu}{4}}\,5^{-\frac{\rho}{2}}
\sum_{n'\ge 7} (\frac{1}{n'})^{\frac98}\, \le \,
\underbrace{\frac{8}{(1- 2^{-\frac{1}{4}})(1-3^{-\frac{1}{4}})
(1-5^{-\frac{1}{2}})}}_{ =:\, K} 
\eqe
so that 
if we choose  $\vep$ sufficiently small to assure
\eq
\label{3rdterm3}
\frac{1}{(n+1)c}\ 2^{5/4}\,K 
\sum_{\nu_1,\nu_2}\frac{(1+2\nu_1)(1+2\nu_2)^{5/4}}{(2\nu_1)!(2\nu_2)!}\ 
\  \vep\  \le \ \frac{1}{30} \frac{1}{(n+2)^{7/8}}\ ,
\eqe
\eqref{3rdterm} is bounded by
\eq
\label{3rdterm2}
\frac{1}{30} \ \frac{1}{(n+2)^{9/8}}\ \vep \ \le \ \frac{1}{30} \ B_\vep(n+2)\ .
\eqe
One may note that the sum appearing 
in \eqref{3rdterm3}  is bounded by 10.

\item{4th term}
\ben 
\label{4thterm}
\begin{split}
&\Biggl|\,\frac{2}{(n+1)c}\,\frac{n-4}{2n}\ a(n)\,+\,
\frac{2}{(n+1)c}\, 
\sum_{{\nu \ge 1 \atop (1+2\nu)n' =n}} 
\frac{n'-4}{2n'}\,\frac{(-1)^{\nu}}{(2\nu+1)!}\ a(n')\,\Biggr|\\
\! \le & \;\frac{\vep}{(n+2)\,c}\,n^{-\frac14}
\underbrace{\left(1\,+\, \frac{3^{\frac14}}{3!}\,+\,\frac{5^{\frac12}}{5!}\,+\,
\frac{7^{\frac98}}{7!}\,+\,\ldots\right)}_{\le \ 5/4}
\,\le\, \frac{1}{30}\ B_\vep(n+2)  \end{split}
\een
for $n$ sufficiently large such that
\ben
(\frac{n+2}{n})^{\frac18}\ n^{-\frac18} \frac{25}{2c} \,\le\, 1\ .
\label{stringent}
\een
\item{5th term}
 \ben
\label{5thterm}
\begin{split}
&\frac{2}{(n+1)c}\,\Biggl|\!
\sum_{{\nu \ge 0 \atop (1+2\nu)n' =n}} \frac{1}{n'}\,
\frac{(-1)^{\nu}}{(2\nu+1)!}\,a(n',1)\,\Biggr| 
\,\le\,
\frac{2}{(n+1)c}\,
\left(\sum_{{\nu \ge 0 \atop (1+2\nu)n' =n}}\frac{n'+1}{n'}\, 
\frac{(2\nu+1)^{1/4}}{(2\nu+1)!}\,\right)\ \frac{\vep^2}{n^{1/4}}\\
&
\,\le\,  \frac{8 }{3c}\, \frac{(n+2)^{9/8}}{(n+1)n^{1/4}}\ 
\frac{\vep^2}{(n+2)^{9/8}}\ 
\ \le \  \frac{1}{30}\ 
\frac{\vep}{(n+2)^{9/8}}\ \le \ \frac{1}{30}\ B_\vep(n+2)\ ,
\end{split}
\een
using that the sum is bounded by $4/3$  and that 
for $\,\vep\,$ sufficiently small  
$\ \frac{8}{3c}\, \frac{(n+2)^{9/8}}{(n+1)n^{1/4}}\,\vep \,\le\, \frac{1}{30}\,$.

\item{6th term}
\ben 
\label{6thterm}
\begin{split}
&\frac{2}{(n+1)c}\,\Biggl|\,
\sum_{{\nu \ge 1 \atop (1+2\nu)n' =n}} 
(-1)^{\nu}\frac{\nu}{(2\nu+1)!}\,a(n')\,\Biggr| 
\,\le\,
\frac{2}{(n+1)c}\,
\sum_{{\nu \ge 1 \atop (1+2\nu)n' =n}} 
\frac{\nu}{(2\nu+1)!}\,B_{\vep}(n')\\
&
\,\le\,  
\frac{2}{(n+1)c}\,
\underbrace{\left(\frac{3^{1/4}}{3!}\,+\,\frac{2\cdot 5^{1/2}}{5!}\,+\,
\frac{3\cdot 7^{9/8}}{7!}\,+\,\ldots\right)}_{\le 3/10}\, B_{\vep}(n)\\
&\,\le\,
\frac{3}{5c}\ \frac{\ (n+2)^{9/8}}{(n+1) n^{1/4}}\, B_{\vep}(n+2)
\,\le\, 
\frac{1}{15}\, B_\vep(n+2)\ \ \mbox{ for }\ \ 
\frac{9}{c} \frac{\ (n+2)^{9/8}}{(n+1) n^{1/4}}\, \le \, 1\ .  
\end{split}
\een
\end{itemize}
We then collect 
\[
0.78\, \frac{n+2}{n+1}\,+\,\frac{1}{10}\,+\,\frac{3}{30}
\  < \ 1\ .
\]
We thus have inductively proven the assertion
\eq
a(n+2) \ \le\  B_\vep(n+2)\ .
\eqe
The most stringent lower bound on $n$ (which equals $\,N_0\,$ for
$\,l\,=\,0$) comes from \eqref{stringent}. This and the other lower 
bounds on $\,n\,$
and upper bounds on $\,\vep\,$ can be relaxed
on distinguishing a number of different cases which we will not do here.\\
It is straightforward to verify the assertion 
for the $\,a(4,l)\,$ by bounding inductively the 
$\mu$-derivatives of \eqref{a4l}.
When taking $\mu$-derivatives of \eqref{ffesin}
we get  
\ben
\label{ffesinder}
\begin{split}
 \!\!  a(n+2,l) 
&\,=\, \frac{1}{n+1}\!\! \!\!\!\!  \sum_{{\nu \ge 1 \atop (1+ 2 \nu)n' =n+2}}\!\!\!\!  
(n'-1) \frac{(-1)^{\nu-1}}{(2\nu)!}\, a(n',l)
\,+\,
\frac{1}{n+1}\!\!\!\!   \sum_{{\nu \ge 1 \atop (1+ 2 \nu)n' =n+2}}\!\!\!\!   
\frac{(n'-1)^2}{n'}\, \frac{(-1)^{\nu-1}}{(2\nu-1)!}\, a(n',l)\\
&\,+\,\frac{1}{(n+1)c}\ 
\sum_{l'+l''=l} {l \choose l'}
\sum_{{\footnotesize \substack{n_1+n_2 = n+2\\ \nu_1,\nu_2\ge 0\\
(1+2 \nu_1) n'=n_1\\ (1+2 \nu_2) n''+n_2}  }}  
(-1)^{\nu_1+\nu_2}\ \frac{(1+2\nu_1)(1+2\nu_2)}{(2\nu_1)!(2\nu_2)!}\,
 a(n',l')\, a(n'',l'')\\
&\,+\,\frac{2}{(n+1)c}\,\frac{n-4}{2n}\,a(n,l)\,+\,
\frac{2}{(n+1)c}\, 
\sum_{{\nu \ge 1 \atop (1+2\nu)n' =n}} 
\frac{n'-4}{2n'}\,\frac{(-1)^{\nu}}{(2\nu+1)!}\,a_(n',l)\\
&\,+\,
\frac{2}{(n+1)c}\,\frac{1}{n}\,a(n,l+1)\,+\,
\frac{2}{(n+1)c}\,\sum_{{\nu \ge 1 \atop (1+2\nu)n' =n}} \frac{1}{n'}\,
\frac{(-1)^{\nu}}{(2\nu+1)!}\,a(n',l+1)\\
&
\,+\,
\frac{2}{(n+1)c}\sum_{\substack{\nu \ge 1\\(1+2\nu)n'=n}}
 (-1)^{\nu}\ \frac{\nu}{(2\nu+1)!}\,\ a(n',l)
\ .
\end{split}
\een
Going from  $\, l\,$ to  $\,l+1\,$, the inductive 
bound for  the l.h.s., i.e. for $\,a(n+2,l)\,$,
is multiplied by a  factor of $\,n+2+l+1\,$.
The respective bounds on the 
linear terms on the r.h.s. take factors of  
\[
1)\ n'+l+1\,,\ \, 
2)\ n'+l+1\,,\ \,
4)\ n+l+1\,,\ \,
5)\ n'+l+1\,,\ \,
6)\ n+l+2\,,\ \,
7)\ n'+l+2\,,\ \,
8)\ n'+l+1\  .
\]
All these factors are strictly smaller than the one for the 
l.h.s. so that the inductive verification of the bound 
for $l >0\,$  follows directly from its verification for $l=0$. 
For the quadratic term (the third term) 
we use the bound \eqref{standard1} which gives
\eq
{l \choose l'}\, \frac{(n'+l')!}{n'!} \,\frac{(n''+l'')!}{n''!}\ \le\
l! \ {n+2+ l \choose n+2}\ .
\label{standard2}
\eqe
The factors $ \frac{(n'+l')!}{n'!} \,$, $\frac{(n''+l'')!}{n''!}\,$
stem from the inductive bounds on the $a(n',l')\,,\ a(n'',l'')\,$.
The expression on the r.h.s. corresponds to the factorials 
appearing in the definition of  $\,B(n+2,l)\,$.  
The sum over $\,l',\,l''\,$ then gives a factor of $\,l+1\,$ which is again
smaller than  $\,n+2+l+1\,$. The remaining part 
of the bound is established as for the third item 
\eqref{3rdterm}. 
\qed

\vspace{.2cm}

We note that (for $\vep'$ sufficiently small) the assumptions
\eqref{a2l} imply the assumptions 
\eqref{f2d}
of Proposition \ref{geom2}
\footnote{Remember that 
$\,a(2,l)\,=\, \pa_\mu^l\, f_2(\mu)|_{\mu=0}\,$.
}. By choosing the signs of 
$\,a(2,l)\,$ appropriately we can also verify the
assumptions of Proposition \ref{geom3}.
As a consequence of Propositions \ref{geom2} 
and \ref{geom3} and of Lemma \ref{boundedaction},
 therefore we have proven  
{\thm 
For  $\,\vep' >0\,$  sufficiently small, the solutions 
of bounded action  \eqref{L0sin} obeying \eqref{a2l} verify
the conditions of Proposition \ref{geom2}. 
For appropriate choices of the $\,a(2,l)\,$ they also verify
 Proposition \ref{geom3}. This implies the  
existence of asymptotically free scalar fields of bounded action
in the mean field limit.
\label{meanfieldasyfree}
}


\subsection{The trivial solution} 
\label{ssec:trivialsol}

It has been proven by Fr\"ohlich~\cite{Froehlich82} and 
Aizenman~\cite{Aizenman82} under mild
assumptions that the pure $\vp^4_4$-theory is trivial,
i.e. interaction free. To be precise the result applies
to the continuum limit of lattice regularised (even) pure 
$\vp^4_4$-theory under the assumption that the theory has infinite
wave function renormalisation. There are no restrictions
on the size of the  $\vp^4_4$ coupling. The result is 
also valid in more than four dimensions without any restriction
on the wave function renormalisation. We note that beyond four dimensions 
it is also known that the critical behaviour of the theory is 
exactly described by the mean field 
approximation~\cite{Ginzburg},~\cite{Froehlich82},~\cite{Aizenman82}. 
The fact that the continuum
limit is interaction free is proven by showing that the truncated
(i.e. connected) four-point function vanishes in this
limit. By inequalities due to Glimm, Jaffe~\cite{GlimmJaffe1974} 
and to Newman~\cite{Newman1975}, the
vanishing of the truncated four-point function implies the vanishing
of the truncated higher $n$-point functions as well. 
The triviality result seems quite robust and has also been confirmed by a comprehensive 
analysis including numerical work~\cite{LueWei1988}. The hypothesis  
on the wave function renormalisation 
has never been verified mathematically. Nor can we shed light on this
hypothesis in our context since we work from the beginning
in the mean field approximation. 

The boundary conditions of pure  $\vp^4_4$-theory
appear to be particularly simple from the point of view 
of Ising type lattice models. On the other hand they appear to be very
special from the point of view of the renormalisation group, where one 
analyses the infinite dynamical system of flow equations 
for the $n$-point functions. 
In fact they correspond
to a fine-tuning problem since it seems unnatural that the 
infinite number of trajectories $\,f_n(\mu)\,,\ n > 4\,,$ all pass 
through $\,0\,$ at the same value of $\mu$, namely at $\,\mu =0\,$. 
This is even more true in the full theory where the boundary 
conditions require that all these trajectories pass through 
 $\,0\,$ at $\,\mu =0\,$ for {\it all} values of the momentum or
position arguments. It will turn out that 
enforcing these conditions tends to make grow higher derivatives of
the $f_n(\mu)$ more rapidly with increasing $\,n\,$ than 
in case of the solutions we have considered so far. 
In any case, since we want to argue that 
our considerations grasp important aspects of scalar 
field theory, it is important to look at the pure  
$\vp^4_4$-theory in our mean field context. 
It will turn out that we can construct the trivial 
solution quite explicitly for all values of the 
renormalisation group parameter and sufficiently small 
bare coupling, thus basically confirming the above cited
results~\cite{Froehlich82},~\cite{Aizenman82}.
   
We start with a bare action 
\ben
L_0(\vp)\,=\, \sum_{n=2,4} \int d^4x \ c_{0,n}\, \vp^n(x) 
\label{purephi}
\een
From \eqref{purephi} we obtain using \eqref{con}, \eqref{A} and \eqref{fA} 
\ben 
f_2(0)\,=\,\al_0\ 2\, (2\pi)^4\, c_{0,2}\ ,\quad
f_4(0)\,=\, 4\,c \ (2\pi)^4\ c_{0,4} \,=\, 4\,\pi^2 \ c_{0,4}\ ,\quad
f_n(0)\,=\,0 \ ,\ \ n > 4\  .
\label{bdpurephi}
\een
As a consequence of the pure  $\vp^4_4$
boundary conditons we have
\begin{lemma}\label{higherterms}
For smooth solutions $\,f_n(\mu)\,$ of \eqref{f42}, \eqref{dyna4}  
respecting the boundary conditions  \eqref{bdpurephi} 
\ben 
\pa_\mu^l \, f_n(0) \,=\, 0 \quad \mbox{ for }\ n \ge 6\ \mbox{ and }\
0 \,\le \,l\, \le\, \frac{n}{2}-3 \ .
\label{nullin}
\een 
\end{lemma}
\noindent
{\it Proof}.
We proceed as usual by induction, in $\, N\,=\, n\,+\,l\,\in \mathbb{N}\,$,
going up in $l$ for fixed $N$ and starting at $\,N=6\,$. 
For  $\,N=6\,$ the assertion just 
corresponds to the boundary condition 

\[ 
f_6(0)\,=\, 0\ . 
\]
For $\,N>6\,$ \eqref{dyna4}, derived $l$ times at $\,\mu=0\,$,
 together with the induction hypothesis implies for $\,l <  \frac{n}{2}-3 $
\[
\pa_\mu^{l+1} f_n(0)\,=\,0
\]
since all other terms appearing in \eqref{dyna3} derived $\,l\,$
times w.r.t. $\,\mu\,$ vanish by induction. We note
in particular that for the products
\[
\pa_\mu^{l_1} f_{n_1}(0)\,\pa_\mu^{l_2} f_{n_2}(0) 
\]
with $\,l_1 +l_2 =l\, $ and $\,n_1+n_2=n+2\,$, 
the condition  $\,l <  \frac{n}{2}-3 \,$ implies
that either $\,l_1 \le   \frac{n_1}{2}-3\, $ 
or $\,l_2 \le   \frac{n_2}{2}-3\, $.   
\qed

\vspace{.2cm}

As a consequence of Lemma \ref{higherterms} we can write smooth solutions verifying
\eqref{bdpurephi} as
\ben 
f_n(\mu)\,=\, \mu^{\frac{n}{2}-2}\ g_n(\mu)\ , \quad n \ge 4\ ,
\label{fg}
\een
where the $\,g_n(\mu)\,$ are smooth.
The system \eqref{dyna4} can then be rewritten as
\ben
\label{dynag}
\begin{split}
\mu^2 g_{n+2}
&\,=\, \frac{1}{n+1}\!  
\sum_{{n_1+n_2 =n+2 \atop n_i \ge 4}}
\! g_{n_1}\,g_{n_2}\\
&\,+\,\mu 
\frac{1}{n+1}\,  g_n\,(2\,f_2\,+\, 1 -\frac{4}{n})
\,+\,\frac{n-4}{n(n+1)}\ g_n\,+\, \frac{2}{n(n+1)}\ \mu \pa_\mu\, g_n\ ,
\quad n \ge 4 \ .
\end{split}
\een
Expanding the $\,g_n\,$ and  $\,f_2\,$ in a (for the moment formal) 
Taylor series around $\,\mu\,=0\,$
\ben
 g_{n}(\mu) \,=\, \sum_{k \ge 0} g_{n,k}\ \mu^k \ ,\quad  
f_{2}(\mu) \,=\, \sum_{k \ge 0} f_{2,k}\ \mu^k 
\label{Taylor}
\een
we find for the coefficients from
\eqref{f42} and \eqref{dyna4} 
\ben
\label{86}
g_{4,k}\,=\,
\frac{1}{3}\,\Bigl((k+1)\,f_{2,k+1}
\,-\,f_{2,k}\,+\, \sum_{0 \le \nu \le k} f_{2,\nu}\, f_{2,k-\nu}\,\Bigr) 
\ \, .
\een
\ben
\label{87}
\begin{split}
g_{n+2,k}
&\,=\,\frac{1}{n+1}\!  
\sum_{{n_1+n_2 =n+2 \atop n_i \ge 4}}
\sum_{_{0 \le \nu\le k+2}} g_{n_1,\nu}\, g_{n_2,k+2-\nu}
\,+\,\frac{2}{n+1} \sum_{0 \le \nu\le k+1} g_{n,\nu}\, f_{2,k+1-\nu}\\   
&\,+\,
\frac{1}{n+1}\,  g_{n,k+1}\,(1 -\frac{4}{n})
\,+\,\frac{n+2k}{n(n+1)}\ g_{n,k+2}\ ,\quad n \ge 4
\end{split}
\een
which can be rewritten  
\ben
\label{88}
f_{2,k+1}\,=\, \frac{1}{k+1}
\left[\,3\, g_{4,k}\,+\,f_{2,k}\,-\, 
\sum_{0 \le \nu \le k} f_{2,\nu}\, f_{2,k-\nu}
\,\right]\ \, , 
\een
\ben
\label{89}
\begin{split}
g_{n,k+2}
\,=\,&
-\,\frac{n-4}{n+2k}\ g_{n,k+1}\,-\, 
\frac{2n}{n+2k}\ \sum_{_{0 \le \nu\le k+1}} g_{n,\nu}\, f_{2,k+1-\nu}\\ 
&\,-\,\frac{n}{n+2k}\ \sum_{{n_1+n_2 =n+2 \atop n_i \ge 4}}
\sum_{_{0 \le \nu\le k+2}} g_{n_1,\nu}\, g_{n_2,k+2-\nu}\,+\,
\frac{n(n+1)}{n+2k}\ g_{n+2,k}
\ \, .
\end{split}
\een
Regularity of the system \eqref{dynag} at $\,\mu\,=0\,$ also implies
for $\,n\,\ge\, 4$
\ben
\label{90}
\begin{split}
&\frac{n-4}{n}\ g_{n,0}
\,+\, 
\sum_{{n_1+n_2 =n+2 \atop n_i \ge 4}}
g_{n_1,0}\  g_{n_2,0} \,=\, 0 \ , \\   
&
\frac{2}{n}\ g_{n,1}\,+\,
\frac{n-4}{n}\ g_{n,1}\,+\,2\! \sum_{{n_1+n_2 =n+2 \atop n_i \ge 4}}
g_{n_1,0}\  g_{n_2,1} \,+\,
g_{n,0}\ ( 2 f_{2,0}\,+\,1-\,\frac{4}{n})
\,=\,0\ .
\end{split}
\een
If we choose freely $\,f_{2,0},\, g_{4,0}\,$, the last two equations 
\eqref{90} fix all other $\,g_{n,0},\,g_{n,1}\,$. 
All terms with $\,f_{2,k}\,$ with  $\,k \ge 1\,$  and $\,g_{n,k}\,$ 
with $\,k \ge 2\,$ are then determined through \eqref{88}, 
\eqref{89}.

\begin{lemma}
\label{g0g1}
We consider smooth solutions $\,f_n(\mu)\,$ of \eqref{f42}, \eqref{dyna4}  
respecting the boundary conditions  \eqref{bdpurephi} and assume that 
\ben
\label{20140} 
 |f_{2,0}| \,\le\,\frac{\,\vep}{4}\ ,\quad  
0 \,\le\, f_{4,0}\,=\, g_{4,0}\,\le\, \frac{\,\vep}{32}
\een
for $\,0 \,<\,\vep\, \le\, 10^{-2}\,$.
Then 
\ben
\label{f210} 
 |f_{2,1}| \,\le\,\frac{\,\vep}{2}\ ,\quad
 |g_{4,1}|  \,\le\,\frac{\,\vep^2}{32}
\een
and for $\,n \ge 6\,$
\ben
\label{gn01}
|g_{n,0}| \,\le\, \frac{\vep^{\frac{n}{2}-1}}{2\,n^2}\ ,\quad 
|g_{n,1}| \,\le\, \frac{\vep^{\frac{n}{2}-1}}{n^2}\  . 
\een 
The constants $\,g_{n,0}\,$ are alternating in sign:
\ben
g_{n,0}\,=\, (-1)^{n/2} \ \left|\,g_{n,0}\,\right| \ .
\een
\end{lemma}

\noindent
{\it Proof}.
For $\,f_{2,1}\,$ we find explicitly from \eqref{88}
\[
f_{2,1}\,=\, 3\,f_{4,0}\,-\, f_{2,0} ( f_{2,0}-1)\,\le\, 
\frac{\,\vep}{2}\ .
\]
Similarly from \eqref{90}
\[
g_{4,1}\,=\, 4 \,g_{4,0}\,f_{2,0}
\quad \mbox{so that}\quad 
\left|g_{4,1} \right| \,\le\,
\frac{\,\vep^2}{32}\ .
\]
We then proceed as usual by induction $\,n\,$, treating first
$\,g_{n,0}\,$. For $\,n \ge 6\, $ we find from \eqref{90}
\[
\left|g_{n,0}\right|\,\le\, 
\frac{n}{n-4}\ \frac14 \sum_{{n_1+n_2 =n+2 \atop n_i \ge 4}} 
\frac{\vep^{\frac{n}{2}-1}}{n_1^2\, (n+2-n_1)^2} 
\,\le\, 
 \frac12  \frac{\vep^{\frac{n}{2}-1}}{n^2} \ .
\]
For  $\,n \ge 6\, $ also
\[
|g_{n,1}|\,\le\,
\frac{2\,n}{n-2}\,\frac12\, 
\sum_{{n_1+n_2 =n+2 \atop n_i \ge 4}} 
\frac{\vep^{\frac{n}{2}-1}}{n_1^2\, (n+2-n_1)^2}
\,+\,
\frac{n}{n-2}\, 
 \frac{\vep^{\frac{n}{2}-1}}{2 \,n^2} \left(\frac{\vep^2}{2} \,+\,1-\frac{4}{n}\right) 
\,\le\, \frac{\vep^{\frac{n}{2}-1}}{n^2}\ . 
\]
The two previous bounds can be verified explicitly for $\,n \le 10\,$.
For $n \ge 12\,$ we use
\[
\sum_{{n_1+n_2 =n+2 \atop n_i\ge 4,\, n_i \in 2\mathbb{N}}} 
\frac{1}{n_1^2\, (n+2-n_1)^2}\,\le\,
\frac{1}{16}
\sum_{{n_1+n_2 =\frac{n}{2}+1 \atop n_i\ge 2, \,n_i \in \mathbb{N}}} 
\frac{1}{n_1^2\, (\frac{n}{2}+1-n_1)^2}
\]
\[
\le\,
\frac{1}{8}
\sum_{{n_1 \le \frac{n}{2}+1 \atop n_1\ge 3}} 
\frac{4}{n_1^2\, (n+2)^2}\,+\, \frac{1}{8} 
\frac{4}{4\, (n-2)^2}
\]
\[
\le\,\frac{1}{2(n+2)^2}\underbrace{(\zeta(2)-\frac{5}{4})}_{\le 1/4}\,+\,  
\frac{1}{8\, (n-2)^2}\,\le\,
\frac{1}{8}\left(\frac{1}{(n+2)^2}\,+\, \frac{1}{(n-2)^2}\right)
\]
and 
\[
\frac{n}{n-2}\,\frac{1}{8}\left(\frac{1}{(n+2)^2}\,+\, \frac{1}{(n-2)^2}\right)
\,\le\, \frac{1}{2\,n^2}\ (1- \vep^2).
\]
The statement on the signs follows
from \eqref{90} by induction in $\,n\,$, 
using that $\,g_{4,0}\,>\,0\,$.
\qed

\vspace{.2cm}

\begin{lemma} 
\label{gnk}
Under the same assumptions as in Lemma \ref{g0g1}
we have the bounds
\ben
\left| g_{n,k}\right| \,\le\,  
2^{k-2} \ \vep^{n/2-1} \  \left(k \,+\, \frac{n-4}{2}\right)! \ \, ,\quad
\left| f_{2,k}\right| \,\le\,  2^{k} \ \vep \ 
\left|k \,-\, 1\right|! \ \, .
\een
\end{lemma}

\noindent
{\it Proof}.
We proceed by induction going up in $\,N\,=\,n\,+\,k\,$ using 
\eqref{89}.
For $\,g_{n,1},\ g_{n,0}\,$ and $\,f_{2,1},\ f_{n,0}\,$ 
we use the bounds from Lemma \ref{g0g1}.
We obtain from \eqref{89}, \eqref{gn01} and Lemma \eqref{gnk}
\ben
\label{bdcoeff}
\begin{split}
\! \left|\,g_{n,k+2}\,\right| 
 \le 2^{k} \, \vep^{\frac{n}{2}-1}   
&\Biggl[
\frac{n-4}{2(n+2k)} \Bigl(k+1 + \frac{n-4}{2}\Bigr)!\,+\, 
\frac{\vep\, n}{n+2k}\! \sum_{_{0 \le \nu\le k+1}} 
\Bigl(\nu + \frac{n-4}{2}\Bigr)!\, 
\left|k-\nu\right|!\\ 
&\,+\, \frac{n}{4(n+2k)} \sum_{{n_1+n_2 =n+2 \atop n_i \ge 4}}
\sum_{_{0 \le \nu\le k+2}} 
\Bigl(\nu \,+\, \frac{n_1-4}{2}\Bigr)!\ \;
\Bigl(k+2-\nu+ \frac{n_2-4}{2}\Bigr)!\\
&\,+\,
\frac{n(n+1)\,\vep}{4(n+2k)}\ \Bigl(k \,+\, \frac{n-2}{2}\Bigr)!\Biggr]
\end{split}
\een
\ben
\begin{split}
&\le 2^{k} \, \vep^{\frac{n}{2}-1} \Bigl(k +2+\frac{n-4}{2}\Bigr)! 
\Biggl[
\frac{n-4}{2n}\, \frac{2}{n}+ 
\frac{2\,\vep\,n}{n}\, \frac{2}{n}
+\frac{n}{4n}\, \frac{n}{2} \, \frac{2}{n}+
\frac{n(n+1)\,\vep}{4n}\,  \frac{2}{n}\,\Biggr]\\
&\le\, 2^{k} \, \vep^{\frac{n}{2}-1} \,  \Bigl(k +2+\frac{n-4}{2}\Bigr)! 
\ \ .
\end{split}
\een
We used
\[
\sum_{0\le \nu \le n-a} (n-\nu)!\, \nu! \, \le \,2\, n! \quad 
\mbox{ for }\  a \in \mathbb{N}\,,\ \  a \le n  
\]
and 
\[
\sum_{0\le \nu \le k} (a+\nu)!\, (b+k-\nu)! \,\le\,  
\sum_{0\le \nu \le k} (A+\nu)!\, (A+k-\nu)! \quad \mbox{with} \quad A=\sup(a,b)\, ,\ \
 a,b \in \mathbb{N}\ .
\]
For $\,n=2\,$   the bound follows from \eqref{88} and Lemma \ref{g0g1}. 
\qed

\vspace{.2cm}

We note that the bounds we derived are not sufficient to prove 
convergence of the Taylor expansion around $\, \mu \,=\,0\,$, 
in contrast to the bounds \eqref{undetbound}.
So \eqref{Taylor} still stand as formal power series. 
We think the factorial behaviour of the  bounds is not far from optimal 
and trace the large size of the derivatives back to the particular 
boundary conditions. We now show that there exist
smooth solutions corresponding to the formal power series \eqref{Taylor}.    

{\prop 
There exist 
smooth solutions $\,f_n(\mu)\,$ of \eqref{f42}, \eqref{dyna4}  
respecting the boundary conditions  \eqref{bdpurephi}.
They vanish in the limit $\,\mu_{max}= \ln\frac{1}{\al_0} \to \infty\,$.
\label{5}
}
\vspace{.2cm}

\noindent
{\it Proof}.
We study two-point functions $\,f_2(\mu)\,$ of the form
\ben
f_2(\mu)\,=\, \sum_{n\ge 1} a_n\ \frac{x_n^{n-1}}{1\,+\, x_n^n}\ , 
\quad\mbox{ where }\quad
x_n\,=\, n\, \mu \quad\mbox{and}\quad\  |a_n| < 1 \ . 
\label{f2tri}
\een
This ansatz is the most important ingredient in our construction
of the trivial solution. 
If it is well-defined, 
then all the $\,f_n(\mu)\,$   and thus all the $\,g_{n}(\mu)\,$ 
are determined  as functions of  $\,f_2(\mu)\,$ as follows from
\eqref{f42}, \eqref{dyna4}.
Expanding as in 
\eqref{Taylor}
\[
f_2(\mu)\,=\, \sum_{k \ge 0} f_{2,k}\ \mu^k\ ,
\]
we find for the Taylor coefficients
\ben 
f_{2,k}\,=\, (k+1)^{k}\  \sum_{\rho = 1}^{k+1}\  a_{\{\frac{k+1}{\rho}\}}\
(-1)^{\rho-1}\ (\,\frac{1}{\rho}\,)^{\,k} \ .
\label{umordnung}
\een
Here we set $\,a_0:=0\,$, \, and for integers $\,n\, , \ m\,$
\ben
\{\frac{n}{m}\} :=\ \Biggl\{ \begin{array}{r@{\quad , \quad}l}  
\frac{n}{m} & \mbox{if }\ \frac{n}{m} \ \in \mathbb{N}\\ 
0 & \mbox{otherwise } 
\end{array} 
\ .
\een
We have in particular
\ben
f_{2,0}\,=\,a_1\ ,\quad f_{2,1}\,=\, 2\, a_2 \,-\, a_1\ . 
\label{a1tri}
\een
Choosing $\,f_{2,0}\,$ and $\,f_{4,0}\,$ such that the 
assumptions of Lemma \ref{g0g1} are fulfilled, 
Lemma \ref{gnk} implies for  smooth solutions of \eqref{f42}, \eqref{dyna4}
respecting the boundary conditions  \eqref{bdpurephi}
\[
\Bigl|\,f_{2,k}\,\Bigr|\,\le \, 2^{k}\ \vep \ |k-1|! \ \  .
\] 
We then claim that the coefficients $\,a_n\,$ 
in \eqref{f2tri} obey the bounds
\ben
\left| a_{n} \right| \,\le\,   4\,(\,\frac{3}{4}\,)^{\,n}\ \vep \ . 
\label{trivex}
\een
The claim is easily verified for $\,a_1\,$ to $\,a_3\,$  
using Lemmata \ref{g0g1} and \ref{gnk}. 
For  $\,n\,\ge\,3\,$ we obtain inductively from  
\eqref{umordnung}\,:
\ben 
\begin{split}
|\, a_{n+1}\,|\,&\le\,
2^n \ \frac{(n-1)!}{(n+1)^{n}} \ \vep   
\,+\, 
 \sum_{\rho = 2}^{n+1}\  \Bigl|\,a_{\{\frac{n+1}{\rho}\}}\Bigr|
\ (\,\frac{1}{\rho}\,)^{\,n} \\
&\le\ 
 \left( (\frac{2}{e})^{\,n}\ \frac{1}{n^2}\,+\, 4\,\sum_{\rho = 2}^{n+1}\ 
 (\,\frac{3}{4}\,)^{\,\frac{n+1}{\rho}}\   (\frac{1}{\rho})^n \right)\ \vep 
\ \le \  4\ (\,\frac{3}{4}\,)^{\,n+1}\ \vep 
\end{split}
\een
using that 
\[
\sum_{\rho = 2}^{n+1}\ 
 (\,\frac{3}{4}\,)^{\,\frac{n+1}{\rho}}\   (\frac{1}{\rho})^n \ \le \
\sum_{\rho = 2}^{n+1}\  (\frac{1}{\rho})^n \,\le\, \zeta(n) -1 \le 2^{1-n}\ .
\]
The bound \eqref{trivex} implies absolute 
convergence of the series in
\eqref{f2tri}, uniformly in $\,\mu\,$ so that $\,f_2(\mu)\,$ is smooth 
and well-defined for $\,0\, \le\,\mu\, \le\,\ln \frac{1}{\al_0}\,$. 
The free choice of $\,f_{2,0}\,$ and $\,f_{4,0}\,$ fixes 
$\,a_1$ and $\,a_2$.
All $\,a_n\,,\  n \ge 3\,$, are uniquely determined by 
\eqref{88}, \eqref{89} and \eqref{90} as a consequence
of the boundary conditions \eqref{bdpurephi} and the smoothness
condition.\\[.1cm] 
Uniform absolute convergence in $\,[0,\infty)\,$
of the series \eqref{f2tri} and its derivatives
\[
\sum_{n\ge 1} a_n\  \pa_\mu^l\ \frac{x_n^{n-1}}{1\,+\, x_n^n} 
\]
and the fact that
\[
 \lim_{\mu \to \infty} \pa_\mu^l\,\frac{x_n^{n-1}}{1\,+\, x_n^n} \,=\, 0
\]
imply 
\ben 
\lim_{\mu \to \infty} \pa_\mu^l\,f_2(\mu) \,=\, 0\ \quad \forall l \ge 0\ .  
\een
The functions  $\,\pa_\mu^l\,f_n(\mu)\,$  for  $n\,\ge\,4\,$ are then determined 
from $\,\pa_\mu^l\,f_2(\mu) \,$ through \eqref{f42}, \eqref{dyna4}. 
Proceeding by induction in $n\,\ge\,4\,$ one finds straightforwardly
\begin{itemize}
\item 
The solutions $\,f_n(\mu)\,$ are smooth bounded functions of $\,\mu\,$.


\item 
Together will all derivatives they have vanishing limits for 
$\mu_{max} \to \infty\,$, i.e. $\,\al_0 \to 0\,$:
\ben  
\lim_{\mu_{max} \to \infty} \pa_\mu^l\, f_n(\mu_{max}) \,=\, 0\ . 
\een
\end{itemize}
\qed

\vspace{.2cm}

We collect the previous findings in the following  
{\thm 
Triviality of weakly coupled mean field pure $\,\vp^4_4$-theory:\\[.1cm]    
For the boundary conditions \eqref{bdpurephi}, setting
\[
0 \,\le \,c_{0,4} \,\le \, \frac{\vep}{2^7\,\pi^2}\ ,\qquad
|c_{0,2}| \,\le \, \,\La_0^{2}\  \frac{\vep}{2^7\,\pi^4} \ \, ,
\qquad
 0 \, < \, \vep \, \le \, 10^{-2}\ ,
\]
the solutions of the mean field flow equations vanish 
in the UV limit $\,\mu_{max} \to \infty\,$,  i.e. on removing the UV cutoff 
$\,\Lao^{-2}\,=\,\al_0\,\to \, 0\,$. 
}    

\vspace{.3cm}

\noindent
We note that the upper limit on $\,\vep\,$ is certainly not optimal
and could be easily improved. A more ambitious project, which  however
does not
seem to be without reach either, is to include also large values 
of $\vep\,$ exceeding $\,1\,$.    

\vspace{.2cm}

\noindent
We close this section with two {\bf general remarks}:

\begin{itemize}
  
\item {\it The Landau pole} 

In perturbative field theory the triviality of pure $\,\vp^4_4$-theory
reflects itself in the so-called Landau pole of the energy dependent 
coupling when going to high energies. This means that when we fix 
the physical coupling at low energies - at $\,\mu\,=\,\mu_{max}\,$ 
in our setting - via 
\ben
g(0)\,:=\, f_2(\mu_{\max})\ , 
\label{renorm}
\een
then 
\[
g(\la) \,:=\, f_2(\mu_{max}-\la)
\] 
diverges at a finite value of $\,\la\,$ unless we let 
$\,g(0)\, \to\, 0\,$ which implies triviality.
 This is indeed the
case for our solution. If we truncate for simplicity 
the expression \eqref{f2tri} at lowest order setting 
\ben
f_2(\mu)\,=\, a_1 \frac{1}{1+\mu}
\label{f2tt}
\een 
we get 
\[
g(\la) \,=\, \frac{g(0)}{1\,-\,\be\, g(0)\,\la}
\quad \mbox{with}\quad
\beta\,=\, \frac{1}{a_1}\ .
\]
The Landau pole is situated at $\la_L\,=\, \frac{1}{\be\,g(0)}\,$. 
In physical perturbation theory one normally chooses  
$\,f_4(\mu_{\max})\,$ to define $\,g(0)\,$, but this does not change
the reasoning since  $\,f_2\,$ and $\,f_4\,$ can be
expressed in terms of each other  and behave in a similar way 
for large $\mu\,$. Nor do the conclusions change
when taking the full expression  \eqref{f2tri} instead of
\eqref{f2tt} since all entries
in the absolutely convergent series in  \eqref{f2tri} 
behave similarly for $\,\mu \to \infty\,$. 
Since the perturbative truncations get out of control way before
the Landau pole divergence occurs, perturbation theory does not
allow to make hard statements about triviality.   

\item {\it Perturbation theory}

The solutions we considered in the previous sections are not perturbative, 
which is reflected in the fact that the bounds
from Propositions \ref{geom2}, \ref{geom} and from section \ref{ssec:solbounded} 
do not involve a power  proportional to $\,n\,$ of  the small 
parameters $\,\de\,$ or $\,\vep\,$. For the trivial
solution the perturbative behaviour 
w.r.t. the bare coupling is revealed  by the factor of
$\,\vep^{n/2-1}\,$ apppearing in the bounds of Lemma \ref{gnk}. 
It should also be possible and would  be interesting
to reexpress the formal power series
in $\,\vep\,$ as formal power series in the renormalized coupling
$\,g(0)\,$ \eqref{renorm} and to show that the coefficients of
these series are termwise finite for $\mu_{max}\,\to\, \infty\,$. 
This would correspond to the perturbative renormalizability proof
for $\vp_4^4$-theory. Our nonperturbative proof implies finiteness
and even triviality  for $\mu_{max}\,\to\, \infty\,$, but we did not 
analyse the expansion in powers of $\,g(0)\,$.

\end{itemize}

\section{The flow equations for the effective potential}
\label{sec:1PI}

The Wilson flow equations for the effective action 
$\, L^{\al_0,\al}(\vp)\,$ can 
be transformed into  flow equations for the effective  
potential, the generating functional of the one-particle
irreducible (1PI) functions, on performing a Legendre 
transformation  \cite{Wetterich}, \cite{Morris}, \cite{KKS}.
We denote the regularised effective potential as
$\,\Ga^{\al_0,\al}(\Phi)\,$. One expects that general results 
for a given field theory which can be deduced from the effective
action, can also be derived from the effective potential. 
Specific properties of the connected and of the 1PI functions 
may of course be different. 
In the mean field approximation the two schemes
are no more strictly equivalent. Nevertheless, 
since we presume that our results on
$\,\vp^4$-theory are generic, we would like to confirm
that the reasoning from section \ref{sec:solutions} can also be applied to 
the 1PI formalism. The analysis becomes more
complicated so that we will restrict ourselves in this paper 
to a result analogous to Propositions \ref{geom2} and \ref{geom3}
in section \ref{ssec:solundet}, namely we will show that 
there exist bounded nontrivial solutions to the flow 
equations which may be asymptotically free. 
A farther reaching analysis of the 1PI functions
is left for the  future.

\vspace{.3cm}

\noindent
The  moments  of the effective potential 
$\,\Ga^{\al_0,\al}(\Phi)\,$ or 1PI $n$-point functions are denoted
as $\ \Ga_n^{\al_0,\al}(p_1,\ldots,p_{n})\ $. They obey the FEs 
~\cite{KKS}
\eq
\partial_{\al} \Ga_n^{\al_0,\al}(p_1,\ldots,p_{n}) \,=\,
\label{fega1}
\eqe
\[
\frac{1}{2}\int_p \frac{1}{1+[\Ga^{\al_0,\al}_2\,C^{\al_0,\al}](p)}\,  
\dot{C}^{\al}(k)\, \hat \Ga_{n+2}^{\al_0,\al}(p,-p,p_1,\ldots,p_{n})\,
\frac{1}{1+[\Ga^{\al_0,\al}_2\,C^{\al_0,\al}](-p)}\ ,
\]
where the functions $\,\hat \Ga_{n+2}^{\al_0,\al}\,$ are given by
\[
\hat \Ga_{n+2}^{\al_0,\al}(p,-p,p_1,\ldots,p_{n})\,=\,
\Ga_{n+2}^{\al_0,\al}(p,-p,p_1,\ldots,p_{n})
\,+\,
\sum_{v=2}^{n/2} \sum^{(v)}_{\{b_j\}} (-1)^{v-1} \,\times\,
\]
\eq
\mathbb{S} \left[\,\prod_{k=1}^{v-1} \left(
\Ga_{b_{k}+2}^{\al_0,\al}(q'_{k-1},p_{i_k+1}, \ldots,p_{i_k+b_k})
C_{\Ga_2}^{\al_0,\al}(q'_{k})\,\right)\,
\Ga_{b_{v}+2}^{\al_0,\al}(q'_{v-1},-p,p_{i_v+1}, \ldots,p_{n-1})\,\right]
\label{fega2}
\eqe
with
\[
q'_0\,=\, p\ ,\ \
q'_k\,=\, p\,+\,\sum_{j=1}^{b_1+\ldots b_k} p_j\, ,\  \ k \ge 1\, ,\quad
b_j \in 2 \mathbb{N}\ , \ \ \sum_{j=1}^{v} b_j\,=\, n\ ,\quad
i_k \,=\,\sum_{j=1}^{k-1} b_j\ .
\] 
Here $\,\sum^{(v)}_{\{b_j\}}\,$ indicates the sum over all partitions
of $\,n\,$ into $v$ packets, the cardinality of the packets being an even integer. 
The symbol $\,\mathbb{S}\,$ has the same meaning as in
\eqref{CAGFEsexpand}.
Note that in the symmetric $\vp_4^4$-theory 
all odd 1PI functions vanish.   
The function $\,C_{\Ga_2}^{\al_0,\al}(p)\,$ denotes the regularised
complete two-point function
\eq
C_{\Ga_2}^{\al_0,\al}(p)\,=\, \frac{C^{\al_0,\al}(p)}{1\,+\, 
\Ga^{\al_0,\al}_2(p,-p)\,C^{\al_0,\al}(p)}\ .
\label{comp2}
\eqe
As compared to ~\cite{KKS} we thus have resummed the two-point function
insertions into \eqref{comp2}.


\subsection{The mean field limit of the 1PI functions}
\label{ssec:meanf1PI}

We want to analyse the mean field limit of 
\eqref{fega2}. The procedure is analogous to that 
of section \ref{sec:meanfield}.
The mean field dynamical system  is obtained from \eqref{fega1}
and \eqref{fega2} 
by replacing the $\, {\Ga}^{\al_0,\al}_{n}(p_1,\ldots p_{n})\,$
by their zero momentum values. 
We call the corresponding mean field  
functions   $\, G^{\al_0,\al}_{n}\,$. 
The flow equations for these functions write for $\,n \in 2\mathbb{N}\,$
\eq
\partial_{\al} G_n^{\al_0,\al} \,=\,
\frac{1}{2}\,
\sum_{v=1}^{n/2}  (-1)^{v-1}\, I^{\al_0,\al}_{v-1} \prod_{k=1}^{v}\,\sum_{\{b_k\}} 
{n \choose b_1 \ldots b_v}\,
G_{b_{k}+2}^{\al_0,\al}\,
\ .
\label{fegag2}
\eqe
Here the integrals $\,I_n^{\al_0,\al}\,$ are defined as
\eq
I_n^{\al_0,\al}\,=\,
\int_p  \dot C^{\al}(p)\, \left[\frac{1}{1\,+\, 
G^{\al_0,\al}_2\,C^{\al_0,\al}(p)}\right]^2\ 
[C_{G_2}^{\al_0,\al}(p)]^{n}
\label{I1pi}
\eqe
with the definition
\eq
C_{G_2}^{\al_0,\al}(p)\,=\, \frac{C^{\al_0,\al}(p)}{1\,+\, 
G^{\al_0,\al}_2\,C^{\al_0,\al}(p)}\ .
\eqe
As in  \eqref{fega2} we sum over all $\,b_j$ such that
\[
b_j \in 2 \mathbb{N}\ , \quad \sum_{j=1}^{v} b_j\,=\, n\  .
\] 
As in section \ref{sec:solutions} we have replaced the mass 
parameter by $0\,$, and will in turn restrict the values 
of $\,\al\,$ to the interval $\,[\,\al_0 \,,1\,]\,$.
To factor out the basic scaling behaviour w.r.t.$\,\al\,$
 we write
\eq
G^{\al_0,\al}_{n}\,=:\,\al^{n/2-2}\,  g^{\al_0,\al}_{n}\ ,
\label{Gg1}
\eqe
which gives the dynamical system
\eq
2\,\al \partial_{\al} \,  g_{n}(\al) \,+\,(n-4)\, g_{n}(\al) \,=\,
\sum_{v=1}^{n/2}  (-1)^{v-1}\, \al^{3-v}\,I^{\al_0,\al}_{v-1} 
\prod_{k=1}^{v}\,\sum_{\{b_k\}} 
{n \choose b_1 \ldots b_v}\,
g_{b_{k}+2}(\al)
\ .
\label{fegagg}
\eqe
We set for $\,v \ge 1$
\[ 
J_{v-1}(\mu)=\al^{3-v}\,I^{\al_0,\al}_{v-1} = 
\int_p \al^{3-v} 
\left[\frac{1}{1+G^{\al_0,\al}_2\,C^{\al_0,\al}(p)}\right]^2\,  \dot C^{\al}(p)\,
\,  [\,C_{G_2}^{\al_0,\al}(p)\,]^{v-1} 
\]
\ben
 \,=\, 
\int_q \,  \dot C^1(q)\, \left[\frac{1}{1\,+\, 
g^{\al_0,\al}_2\,C^{\,\ga,1}(q)}\right]^2\, 
[\,C_{g_2}^{\,\ga,1}(q)\,]^{v-1} \ ,
\label{jv1}
\een
where $\,0\,\le\,\ga\,=\,\frac{\al_0}{\al}\,\le\, 1\,$
and $\, q^2 =\al \,p^2\,$. 
Passing to the variable $\,\mu\,=\, \ln \frac{\al}{\al_0}\,=\, 
\ln\frac{1}{\ga}\, \,\ge\, 0\,$ and 
writing in shorthand $\,g_n(\mu)\,:=\,g_n^{\al_0,\al}\, $ we then get
\eq
2\,\partial_{\mu} \,  g_{n}(\mu) \,+\,(n-4)\, g_{n}(\mu) \,=\,
\sum_{v=1}^n  (-1)^{v-1}\, J_{v-1}(\mu)\,\prod_{k=1}^{v}\,\sum_{\{b_k\}} 
{n \choose b_1 \ldots b_v}\,
g_{b_{k}+2}(\mu)
\ .
\label{fegamu}
\eqe
Setting finally
\eq
g_n(\mu)\,=:\, (n-2)! \  h_n(\mu) \,, 
\, \mbox{ which implies }\ \,
h_n(\mu) \,:=\, \al^{2-\frac{n}{2}} \ \frac{1}{(n-2)!}\ 
G^{\al_0,\al}_n\ , 
\label{hn}
\eqe
we obtain the mean field 1PI flow equations in the form 
suited for our subsequent analysis
\[
 J_{0}(\mu)\,h_{n+2}(\mu)
\]
\vspace{-.75cm}
\ben
\qquad =\,
\sum_{v=2}^{n/2}  (-1)^{v}\, J_{v-1}(\mu)\,\sum_{\{b_k\}} \prod_{k=1}^{v}\,
h_{b_{k}+2}(\mu)\,+\, \frac{2}{n(n-1)}\partial_{\mu} \,  h_{n}(\mu)\,+\,
\frac{n-4}{n(n-1)}\, h_{n}(\mu) 
\ .
\label{hamu}
\een
This system of equations is obviously more complicated than
the system \eqref{f42}, \eqref{dyna4} due to the appearance
of the functions $\,J_\nu\,$. These functions are well-behaved
and straightforward to control, as will be seen in the next section.
We shall be able to make analogous statements to 
Propositions \ref{geom2} and \ref{geom3} for the 1PI functions.
For farther reaching results corresponding to Proposition \ref{geom}
and section \ref{ssec:solbounded},
we would need even sharper bounds on arbitrary
derivatives of these functions.


\subsection{Solutions for the 1PI case}
\label{ssec:sol1PI}

We establish  bounds on smooth solutions of the dynamical 
system \eqref{hamu}. The bounds are expressed in terms of positive constants 
$\, K\, ,\ \de \, , \ \be\,$, and a smooth positive function 
$\,\de(\mu)\,$ satisfying
 $\,\de(\mu)\,\le\,\de\,$. Our assumptions are similar
as in  \eqref{f2d}, \eqref{f2asy1}.
We assume $\,K\,$ to be sufficiently large and
$\,\de,\, \be\,$ to be sufficiently small. The bounds will turn out to hold
for 
\eq
K \, \ge \, \frac{4}{c}\ ,\quad  
\be \,\le \, 2\  ,\quad \de \,\le \, \frac{c}{4}  \ ,
\quad \mbox{with (as before)}\quad  
c\,=\,\frac{1}{16\pi^2}\ .
\label{assu1}
\eqe
We assume the following properties of the
two-point function $\,h_2(\mu)\,$: 
 \eq
 h_2(\mu)\,=\, -\,\de(\mu)\ ,\quad
\pa_\mu \de(\mu) \,=\, \be \, \de^2(\mu)\ ,\quad
0 < \de(\mu) \le \de \quad \forall \mu \in [\,0\,,\,\ln\frac{1}{\al_0}\,]\ .
\label{assu2} 
\eqe
From the assumptions \eqref{assu2} it follows that
\eq
\pa^{\,l}_\mu\, \de(\mu)\,=\,\be^l\ \de^{l+1}(\mu)\ l\,!\ \ ,\quad
\pa^{\,l}_\mu\, \de^N(\mu)\,=\,\be^l\ \de^{l+N}(\mu)\ \frac{(N+l)\,!}{N\,!}\ \ .
\label{deabl}
\eqe
It is possible to make assumptions more general than
\eqref{assu1}, \eqref{assu2},
for example $\,\pa_\mu \de(\mu) \,=\, \be(\de(\mu))\,$, with 
 $\be(x)\,=\,O(x^2)\,$, and $\be\,$ analytic 
in a disc of sufficiently large radius, typically 
larger than $\,1/\de\,$.

{\lemma 
\label{jv}
Under the assumptions \eqref{assu1}, \eqref{assu2}
\eq
\frac{c}{(1\,+\,\delta(\mu))^2}\,
 \,\le \, J_0(\mu)\,\le\,\frac{c}{(1\,-\,\delta(\mu))^2}\,
 \ ,
\eqe
and for $\,v \in \mathbb{N}\ \,$
\eq
0\, <\, J_v(\mu) \, \le \, \frac{c\ (1-\ga)^v}{\ (1-\de(\mu))^{2+v}}\
\ \, . 
\eqe
}
\vspace{.1cm}

\noindent
{\it Proof}.
The first bound is immediate from the definition
\eqref{jv1} and since
\ben
C^{\ga,1}\,=\, \int_{\ga}^1 d\zeta\ e^{-\zeta\,q^2}\,\le \,1 \ .
\een
Regarding the second bound
we write
\ben
\begin{split}
J_{v}(\mu)\,=\,& 
\int_q \,  \dot C^1(q)\, \left[\frac{1}{1\,+\, 
h_2(\mu)\,C^{\,\ga,1}(q)}\right]^2
\, 
[\,C_{h_2}^{\,\ga,1}(q)\,]^{v} \\
\,=\,& \int_{\ga}^1 d\ga_1 \ldots d\ga_v\,
\int_q e^{-(1+\ga_1+ \ldots +\ga_v)\,q^2}\  
\left[\frac{1}{1\,+\, 
h_2(\mu)\,C^{\,\ga,1}(q)}\right]^{2+v}\\
\,\le\,&
\,\frac{c}{(1-\de(\mu))^{2+v}}\
\int_{\ga}^1 d\ga_1 \ldots d\ga_v\,
\frac{1}{(1+\ga_1+ \ldots +\ga_v)^2}
\ \le\
\,\frac{c\ (1-\ga)^{v} }{\ (1-\de(\mu))^{2+v}}\ \, .
\label{jvbd}
\end{split}
\een
\qed

\vspace{.2cm}

{\lemma 
\label{logder}
For $\,n \in \mathbb{N}\,$
\eq
\left(x \frac{d}{dx}\right)^n\,=\,
\sum_{\nu=1}^n a(n,\nu) \,x^\nu\,\frac{d^\nu}{dx^\nu} \ .
\label{expder}
\eqe
The positive integers $\,a(n,\nu) \,$ satisfy
\eq
 a(1,1)\,=\,1\ ,\quad  a(n,n)\,=\, a(n,1)\,=\,1 \  .
\label{init}
\eqe
\eq
 a(n+1,\nu) \,=\, a(n,\nu)\, \nu\,+\, a(n,\nu-1)\ ,\quad 2\le \nu \le n\ ,
\label{recur}
\eqe
\eq
a(n,\nu) \,\le \, 2^n\, \frac{n!}{\nu!}\ .
\label{annbound}
\eqe
}
{\it Proof}.
It is obvious that the expansion \eqref{expder} holds 
with nonnegative integer coefficients. 
The relations \eqref{init} are also immediate. Furthermore
\[
x\,\frac{d}{dx}\sum_{\nu=1}^n a(n,\nu) \,x^\nu\frac{d^\nu}{dx^\nu} 
\,=\,\sum_{\nu=1}^n \left[\,a(n,\nu)\, \nu \,x^\nu\frac{d^\nu}{dx^\nu}\,+\,
a(n,\nu)\, \,x^{\nu+1}\frac{d^{\nu+1}}{dx^{\nu+1}}\,\right] \ ,
\] 
which gives \eqref{recur}\,;\,
\eqref{annbound} (which is not optimal)
follows directly by induction on $\,n+\nu \,\ge\, 2\,$.
\qed

\vspace{.2cm}

\noindent
Lemma \ref{logder} says that for $\,l \ge 1\,$
\eq
\pa_\mu^l \,=\, (-\ga \pa_\ga)^l\,=\, (-1)^l \sum_{\la=1}^l 
 a(l,\la)\,\ga^\la\,\frac{\pa^\la}{\pa\ga^\la}\ . 
\label{muga}
\eqe
We set for $\ 0 \le \ga \le 1\,$
\eq
I_N(\ga) \,:=\, \int_\ga^1 d\ga_1\ldots d\ga_N\,
 \frac{1}{(1+\ga_1+\ldots+ \ga_N)^2} \ .
\eqe
{\lemma 
\label{INga}
\ben
\mbox{For} \ \ l\, \ge\, 1\,\quad |\,\pa_\mu^l\,I_{N}(\ga)\,| 
\ \le\   \ga\ 2^{N+2l+2}\,(l+1)\,!\ \, .
\een
}
{\it Proof}.
We have
\[
\!\pa_\ga^l\,I_{N}(\ga) \,=\, 
\sum_{\la=1}^l {l \choose \la}\, (-1)^l\,\frac{N!}{(N-\la)!}\,
\int_\ga^1 d\ga_{\la+1}\ldots d\ga_N\, \frac{ (l-\la+1)!}{(1+\la \ga+ \ga_{\la+1}
+\ldots+ \ga_N)^{2+l-\la}}
\]
and
\ben
\nonumber
\begin{split}
\int_\ga^1 \frac{d\ga_{1+\la}\ldots d\ga_N\, }{(1+\la \ga+ \ga_{\la+1}
+\ldots+ \ga_N)^{2+l-\la}}\,\le\,&
\sum_{j=0}^{N-\la} {N-\la \choose j} 
2^{-(N-\la)}\frac{1}{(1+\frac{j}{2})^{2+l-\la}}\\ 
\le\,&
\ 2^{2+l- N}\,\sum_{j=0}^{N-\la} {N-\la \choose j}\frac{1}{(2+j)^{2+l-\la}} 
\ \, .
\end{split}
\een
To obtain the last bound we split each integration interval 
$\,[\ga,\,1]\,$ into its lower and upper half segment and bound the integrand
for each choice of segments by its $\,\sup\,$. Then
\ben
\begin{split}
|\,\pa_\ga^l\,I_{N}(\ga)\,| \,\le & \ 2^{2+l- N}
\sum_{\la=1}^l {l \choose \la}\,\frac{N!}{(N-\la)!}\, 
 (l-\la+1)!\sum_{j=0}^{N-\la} {N-\la \choose j}\frac{1}{(2+j)^{2+l-\la}}\\ 
\le\,& \sum_{\la=1}^l\sum_{j=0}^N \, 2^{2+l-N+l+2N-\la} (l-\la+1)!\ \,\la!\
\ \frac{1}{(2+j)^{2+l-\la}}\\ 
\le&
\  2^{2+2l+N}\ l!\  \sum_{\la=1}^l   2^{-\la} \ (\zeta(2+l-\la)\,-\,1)\\
\le&\ \  2^{N+l+1}\ l!\ l  \ ,
\label{ing}
\end{split}
\een
where we used the well-known bounds for the $\,\zeta$-function
\[
\zeta(n)-1 \,\le\,2^{1-n}\ \mbox{ for } \ n \,\ge\, 2\ .
\]
From \eqref{muga}, Lemma \ref{logder} and  \eqref{ing} 
we then get 
\[
|\,\pa_\mu^l\,I_{N}(\ga)\,| 
\ \le\ 2^{N+1}
\sum_{\la=1}^l 2^l\  \frac{l!}{\la\,!}\ 2^\la\  \la!\ \la\ \ga^{\,\la}\,
\]
\ben
\,\le\,
 2^{N+l+1}\,l\,!\ \sum_{\la=1}^l \la\ (2\ga)^\la\,\,\le\,
2^{N+2l+2}\  \ga\ (l+1)\,!\ \, .
\label{linga}
\een
\qed

\vspace{.2cm}

\noindent
The bound of Lemma \ref{INga} can be improved if
$\,\ga\,$ is close to 1.

\vspace{.2cm}

\begin{lemma} 
\label{jlv}
Under the assumptions \eqref{assu1}, \eqref{assu2} 
and for $ l \,\ge \,1 \,$
\ben
\left|\, \pa_\mu^l \,  J_v(\mu)\,\right| \, \le \, 
\,\frac{\ga}{\,\pi^2}\ 2^{2v+3l}\ (l+1)!\     
 \ ,\quad
\left|\, \pa_\mu^l   J_0(\mu)\,\right| \, \le \, 
\de(\mu) \, \frac{\ga\,}{2\, \pi^2}\,2^{3l}\,(l+1)!\ \ . 
\een
\end{lemma}
{\it Proof}.
Setting $\,f(\mu)\,=\,\de(\mu)\, C^{\ga,1}(q)\,$
and expanding, we obtain for $\,v\ge 1$
\[
\frac{1}{[1\,-\,f(\mu)]^{2+v}}\,
=\, 
\sum_{N=0}^\infty \sum_{_{n_1+\ldots n_{2+v}\,=\,N}}\, f(\mu)^N 
\,=\,
\sum_{N=0}^\infty {N+v+1 \choose N}\,
\de^{N}(\mu)\, [\,C^{\ga,1}(q)\,]^N\ .
\]
Using this expression in \eqref{jv1}
\ben
\begin{split}
J_{v}(\mu)\,=\,& 
\int_q \,  \dot C^1(q)\, \left[\frac{1}{1\,-\, 
\de(\mu)\,C^{\,\ga,1}(q)}\right]^2
\, 
[\,C_{h_2}^{\,\ga,1}(q)\,]^{v} \\
\,=\,& \int_{\ga}^1 d\ga_1 \ldots d\ga_v\,
\int_q e^{-(1+\ga_1+ \ldots +\ga_v)\,q^2}\  
\left[\frac{1}{1\,-\, 
\de(\mu)\,C^{\,\ga,1}(q)}\right]^{2+v} \\
\,=\,&
c\,\sum_{N=0}^\infty {N+v+1 \choose N}\,\de^{N}(\mu)\, 
\,\int_{\ga}^1 d\ga_1 \ldots d\ga_{v+N}\,
\frac{1}{(1+\ga_1+ \ldots +\ga_{v+N})^2}\ ,
\label{jvbd0}
\end{split}
\een
and deriving $l$ times w.r.t. $\mu$, using \eqref{deabl} and Lemma \ref{INga},
we then get the bounds
\ben
\label{jvlb}
\begin{split}
\!\!\!\!\!\!\!\!\!| \pa_\mu^l   J_v(\mu)|  \le &\, 
c\,\sum_{N=0}^\infty  {\mbox {${N+v+1 \choose N}$}}
\sum_{\la=0}^l 
 {\mbox {$ {l \choose \la}$}}
\,\be^\la\ \de^{N+\la}(\mu)\,  {\mbox {$\frac{(N+\la)\,!}{N!}$}}\,
\ga \ 2^{N+\nu+2(l-\la)+2}\, (l-\la+1)!\\
\, \le &\ c\,\ga\,\sum_{N=0}^\infty 2^{2N+2v+3l+3}\ \de^{N}(\mu)\sum_{\la=0}^l 
(\be\, \de(\mu))^{\la}\ 2^{-2\la}\, (l-\la+1)!\\ 
\,\le\,& \ c\,\ga\ 2^{2v+3l+3}\ \frac{1}{1-4\,\de(\mu)}\ 
\frac{1}{1-\,\frac14\,\be\de(\mu)}\ 
 (l+1)\,!\ \le\ \frac{\,\ga}{\,\pi^2}\ 2^{2v+3l}\ (l+1)! \ \ . 
\end{split}
\een
In the case $\,v=0\,,\ l\ge 1\,$ we get 
\ben
\label{jolb}
\begin{split}
\!\!\!\!\!\!\! | \pa_\mu^l   J_0(\mu)| \le &
\,c \sum_{_{N=0}}^\infty\! {\mbox {${N+1 \choose N}$}}
\sum_{\la=1}^l  {\mbox{${l \choose \la}$}}
\,\be^\la\, \de^{N+\la}(\mu)\, {\mbox {$\frac{(N+\la)!}{N!}$}}
\ga \ 2^{N+2(l-\la)+2}\, (l-\la+1)\,!\\
\, \le &\ c\,\ga\,\sum_{_{N=0}}^\infty 2^{2N+3l+3}\ \de^{N}(\mu)\sum_{\la=1}^l 
(\be\, \de(\mu))^{\la}\ 2^{-2\la}
 (l-\la+1)!  \\\,\le\,& \ c\,\ga\ 2^{3l+3}\ \frac{1}{1-4\,\de(\mu)}\ 
\frac{\frac14\,\be\,\de(\mu)}{1-\,\frac14\,\be\de(\mu)}\ 
 (l+1)\,!\ \le\ \,\de(\mu)\ \frac{\ga}{2\,\pi^2}\ 2^{3l}\ (l+1)!\ \ . 
\end{split}
\een
We can make explicit a factor of $\,\frac14\,\be\,\de(\mu)\,$ 
in the last line, since  
there is no contribution with $\,\la=0\,$ in the second sum in
\eqref{jolb} as compared to \eqref{jvlb}. 
\qed

\vspace{.2cm}

{\lemma 
\label{prodhl}
\eq
\label{vlbound}
\sum_{\{l_j\}} \prod_{j=1}^{v}\,
\frac{1}{(l_j+2)(l_j+1)}\,\le\,
\frac{(3/2)^v}{(l+2)(l+1)}\, , \quad \mbox{where }\ v\ge 1\, ,\quad 
 l,\, l_j \in \mathbb{N}\, ,\quad 
\sum_{j=1}^v l_j \,=\, l\ . 
\eqe
}
\vspace{.1cm}
\noindent
{\it Proof}.
The statement is evident for $\,v\,=\, 1\,$. 
For  $\,v\,=\, 2\,$ 
we have
\ben
\label{lprod}
\nonumber
\begin{split}
&\ \sum_{l_1 = 0}^l \frac{1}{(l_1+2)(l_1+1)}\,\frac{1}{(l-l_1+2)(l-l_1+1)}
=\sum_{l_1 =0 }^l (\frac{1}{l_1+1}- \frac{1}{l_1+2})\,
 (\frac{1}{l-l_1+1}- \frac{1}{l-l_1+2})\\
&=\,
\sum_{l_1 =0 }^l\Bigl[\, \frac{1}{l+2}\,(\frac{1}{l_1+1}\,+\, \frac{1}{l-l_1+1})\,
\,-\,
\frac{1}{l+3}\,(\frac{1}{l_1+2}\,+\, \frac{1}{l-l_1+1})\\
&\qquad -\,
\frac{1}{l+3}\,(\frac{1}{l_1+1}\,+\, \frac{1}{l-l_1+2})\,
\,+\,
\frac{1}{l+4}\,(\frac{1}{l_1+2}\,+\, \frac{1}{l-l_1+2})\,\Bigr]\\
&=\
2\,(\frac{1}{l+2}\,-\, \frac{1}{l+3}) \,  \sum_{l_1 =0 }^l\frac{1}{l_1+1}\,
\,-\,
2\,(\frac{1}{l+3}\,-\, \frac{1}{l+4}) \, \sum_{l_1 =0 }^l\frac{1}{l_1+2}\\ 
&=\
\frac{1}{(l+2)(l+3)} \,(1-\frac{2}{l+2})\,+\,
\bigl[\frac{1}{(l+2)(l+3)} \,-\, \frac{1}{(l+3)(l+4)}\bigr]
\sum_{l_1 =0 }^l\frac{2}{l_1+2}\\
&\le\
\frac{1}{(l+2)(l+1)}\,
\Bigl[\,\frac{l}{l+3}\,+\,\frac{4(l+1)}{(l+3)(l+4)}\ln(l+1)\,\Bigr]\,\le\,
\frac32\ \frac{1}{(l+1)(l+2)}\  \, .
\end{split}
\een
The 3$^{rd}$ and 4$^{th}$ lines are obtained by expanding the products
and using $\frac{1}{ab}\,=\,(\frac{1}{a}\,+\,\frac{1}{b})\,\frac{1}{a+b}\,$.
The expression in square brackets in the last line reaches its maximal value
$\,1.446$ for $\,l=10\,$.\\ 
We then get by induction $\,v-1 \to v\,\ge 3\,$ 
\ben
\label{lprodind}
\begin{split}
&\quad\ \sum_{\{l_j\}} \prod_{j=1}^{v}\,
\frac{1}{(l_{j}+2)(l_{j}+1)}
\,\le\,
\sum_{_{1 \le l_v \le l-v-1}} 
(\frac32)^{v-1}\ \frac{1}{(l-l_v+2)(l-l_v+1)}\ \frac{1}{(l_{v}+2)(l_{v}+1)}\\
\le\,&
\sum_{_{1 \le l_v \le l-1}} 
(\frac32)^{v-1}\ \frac{1}{(l-l_v+2)(l-l_v+1)}\ \frac{1}{(l_{v}+2)(l_{v}+1)}
\,\le\,
(\frac32)^{v}\ \frac{1}{(l+2)(l+1)}\ ,
\end{split}
\een
where we  applied the previous bound \eqref{lprod} again.
\qed

\vspace{.2cm}

As a consequence of of Lemma \ref{prodhl} we obtain
immediately

{\lemma 
\label{prodhbl}
\ben
\label{jvlbound}
\sum_{\{b_k\}} \prod_{k=1}^{v}\,
\frac{1}{(b_{k}+2)(b_{k}+1)}\,\le\,
\frac{(3/2)^v}{(n+2)(n+1)}\, , \ \mbox{ where }\ v\ge 2\, ,\ 
  b_k \in 2\mathbb{N}\, ,\ 
\sum_{k=1}^v b_k \,=\, n\ . 
\een
}

{\prop 
\label{1pI4}
Under the assumptions \eqref{assu1}, \eqref{assu2} 
\ben
h_4(\mu) \,=\,J_0^{-1}\,\left(\de(\mu)\,-\, \be\, \de^2(\mu)\right) 
\,>\, \frac{(1-\de(\mu))^2}{c}\,\left( \de(\mu)\,-\,\be\,\de^2(\mu)\right)
\, >\,0 \ \, .
\label{h4}
\een
We set 
\ben
 {\cal B}(n,l;\mu)\,:=\,
\frac{\de^{2}(\mu)\,K^{n+l-2}\,}{(n+2)(n+1)\, (l+2)(l+1)}\ \,(n+l-2)! \ . 
\label{bnlm}
\een
For $\,n \,=\, 4\,$, $\,l \,\ge\, 1\,$ and for 
 $\,n \,\ge\, 6\,$ we have the bounds
\ben
\left|\,\pa_\mu^{\,l}\, h_n(\mu)\,\right| \, \le \, {\cal B}(n,l;\mu)\ , \quad
\left|\,h_4(\mu) \,\right| \, \le \,  {\cal B}(4,0)\ \frac{4c}{\de(\mu)} \ .
\label{hmun}
\een
}
\vspace{.1cm}
{\it Proof}.
For the four-point function the dynamical system gives 
\[
\pa_\mu^{\,l}\,[ J_0(\mu) \, h_4(\mu)]\,=\, 
\partial^{\,l+1}_{\mu} \,  h_{2}(\mu)\,-\,
\partial^{\,l}_{\mu} h_{2}(\mu)\,=\,\be^{l}\ \de^{l+1}(\mu)\ l\,!\ \,
-\,\be^{l+1}\ \de^{l+2}(\mu)\ (l+1)\,! \ .
\]
Then \eqref{h4} directly follows from \eqref{assu1} and Lemma \ref{jv}. 
For $\, n\,=\,4\,,\ l \,\ge\, 1\,$ we then get by induction in $\,l\,$
\ben
\begin{split}
&\!\!\!\!\!\!\!\!\left| J_0 \, \pa_\mu^{\,l}\,h_4(\mu)\right| \le \,
\be^{l}\ \de^{l+1}(\mu)\ l\,!\ +\ \be^{l+1}\ \de^{l+2}(\mu)\  (l+1)!
 \\
 +&\ 
\de(\mu)\ \frac{1}{2\,\pi^2}\sum_{\la=1}^l  {l \choose \la} \,
2^{3\la}\ (\la+1)!\ K^{l-\la+2}\ \sup(4c, \de(\mu)) \ \de(\mu)\ \frac{(l-\la)!}{60}\\
\le&\  
\be^{l} \de^{l+1}(\mu)\,  l\,! \,+\,
\be^{l+1} \de^{l+2}(\mu)\,  (l+1)! \,+\,
4c\,\de^2(\mu) \frac{K^{2+l} l! }{120\,\pi^2}
\underbrace{\sum_{\la=1}^l (\frac{8}{K})^{\la} (\la+1)}_{\le 24/K}\  .
\end{split}
\een
The factor $\,[\,\sup\{4c,\de(\mu)\}]^v\,=\, (4c)^v\,$
is due to the fact that underived four-point functions allow for an
additional factor of  $\,4c\,$ in the bound, whereas derived 
functions allow for an additional factor of $\,\de\,$, as follows 
from \eqref{bnlm} and \eqref{hmun}. 
Then the assertion is true if 
\[
\frac{(1+\de^2(\mu))}{c}\left( \be^{l}\ \de^{l-1}(\mu)
\,+\,\be^{l+1}\ \de^{l}(\mu) \,(l+1) 
\,+\, \frac{4c\,K}{5\,\pi^2}\  
K^l \right)\,\le\, \frac{\ K^{2+l}}{30}
\] 
which holds due to \eqref{assu1}.\\[.1cm]
We then proceed by induction in $\,n+l\,$  for   $\,n\,\ge 4\,$.
We apply the induction hypothesis to \eqref{hamu}, 
derived $\,l\,$ times w.r.t. $\,\mu\,$ which can be written
as 
\ben
\begin{split}
 J_{0}(\mu)\,\pa_\mu^l\,h_{n+2}(\mu)
=\,
-&\,\sum_{\la=0}^{l-1}\ { l \choose \la}\
 (\pa_\mu^{l-\la}  J_{0}(\mu))\ \pa_\mu^{\la}\,h_{n+2}(\mu)\\
+&\,\sum_{v=2}^{n/2}  (-1)^{v}\, 
\sum_{\la=0}^{l}\  { l \choose \la}\ \pa_\mu^{l-\la}\,J_{v-1}(\mu))\,\sum_{\{b_k\}}
\Bigl(\pa_\mu^{\la} \prod_{k=1}^{v}\,
h_{b_{k}+2}(\mu)\Bigr)\\ 
+&\, \frac{2}{n(n-1)}\partial^{l+1}_{\mu} \,  h_{n}(\mu)
\,+\,
\frac{n-4}{n(n-1)}\, \partial^{l}_{\mu} \, h_{n}(\mu) 
\ .
\label{hamuder}
\end{split}
\een
We now bound successively the contributions from
the four terms on the r.h.s. of this equation.
\begin{itemize}

\item{The 1st term is bounded by}
\ben
\label{1stterm1pi}
\begin{split}
&\left|\,\sum_{\la=0}^{l-1}\ { l \choose \la}\
 (\pa_\mu^{l-\la}  J_{0}(\mu))\ \pa_\mu^{\la}\,h_{n+2}(\mu)\,\right|\\
\le&\, \sum_{\la=0}^{l-1} { l \choose \la}\
 \frac{\de(\mu)}{2\pi^2}\ 2^{3(l-\la)}\ (l-\la+1)!  
\frac{\de^{2}(\mu)\,K^{n+\la}\,(n+\la)!}{(n+4)(n+3)\, (l+2)(l+1)}\\
\le&\,  \frac{\de(\mu)}{2\pi^2}\, 
\frac{\de^{2}(\mu)\,K^{n}\ 8^{l}}{(n+4)(n+3)(l+2)(l+1)}
\underbrace{\sum_{\la=0}^{l-1}\,  (\frac{K}{8})^{\la}\, 
(n+\la)!}_{\le\, (K/8)^{\,l}\, (n+l)!}\, \le\,   
\frac{\de(\mu)}{2\pi^2} \, {\cal B}(n+2,l;\mu)\ .
\end{split} 
\een  

\item{The 2nd term is bounded by}
\ben
\label{2ndterm1pi}
\begin{split}
&\left|\,\sum_{v=2}^{n/2}  (-1)^{v}\, 
\sum_{\la=0}^{l}\  { l \choose \la}\ 
\pa_\mu^{l-\la}\,J_{v-1}(\mu))\,\sum_{\{b_k\}}
\Bigl(\pa_\mu^{\la} \prod_{k=1}^{v}\,
h_{b_{k}+2}(\mu)\Bigr)\,\right|\\ 
\le& \ \sum_{v=2}^{n/2}  \sum_{\la=0}^{l} { l \choose \la}\
\frac{1}{\pi^2}\ 2^{2(\nu-1)+3(l-\la)} \,(l-\la+1)!\   K^{n+\la}\     \\
&\times\ 
\sum_{\{b_k\}, \{\la_k\}}
 { \la \choose \la_1 \ldots \la_v}\,  \prod_{k=1}^{v}\,
\frac{K^{b_k+\la_k}\ \de^{v}(\mu)\, [\,\sup\{4c,\,\de(\mu)\}\,]^v}
{(b_k+4)(b_k+3)\, (\la_k+2)(\la_k+1)}\ \,(b_k+\la_k)!\\
\le& \ \sum_{v=2}^{n/2}  \sum_{\la=0}^{l} 
\frac{1}{\pi^2}\ 2^{2(\nu-1)+3(l-\la)} \, l!\ (l-\la+1)\,  K^{n+\la}\  
\frac{(n+\la)!}{\la!}  \ (4c\,\de(\mu))^{v} \\
&\times\  
\ \frac{(3/2)^v}{(\la+2)(\la+1)}\ \frac{(3/2)^v}{(n+4)(n+3)}\   
\\
\le& \ \frac{1}{4 \pi^2}\ \frac{K^n\ 8^{l}\ l!}{\,(n+4)(n+3)}\  
\underbrace{\sum_{v=2}^{n/2} [36\,c\, \de(\mu) \,]^\nu}_{\le\, 2 (36\,c\,\de(\mu))^2}  
\underbrace{\sum_{\la=0}^{l}
 \frac{(n+\la)!\  (l-\la+1)\ (K/8)^\la}{(\la+2)!}}_{\le\ 2\, (K/8)^l\, (n+l)!/l!}
\\
\le& \  \frac{1}{\,\pi^2}\ \frac{(36\,c\, \de(\mu))^2\ K^{n+l}\ }{\,(n+4)(n+3) (l+2)(l+1)}\ 
 (n+l)! \ \le \ 
\ \frac{(36\,c)^2}{\,\pi^2}\ {\cal B}(n+2,l;\mu) \ .
\end{split} 
\een  
We have used 
Lemmata \ref{jlv},
 \ref{prodhl} and \ref{prodhbl} 
to sum over the $\,b_k\,$ and over the $\, \la_k\,,\ \sum \la_k = \la\,$.
 Furthermore
we used
\[
\prod_{k=1}^{v}\,\frac{(b_k+\la_k)!}{\la_k!}\ \le\  \frac{(n+\la)!}{\la!}\ .
\] 
\item{The 3rd term is bounded by}
\ben
\begin{split}
\left|\,\frac{2}{n(n-1)}\partial^{l+1}_{\mu} \,  h_{n}(\mu)\,\right|
\ \le\ \frac{2}{n(n-1)}\ {\cal B}(n,l+1) \ \le \
\frac{1}{20\,K}\ {\cal B}(n+2,l) \ .
\label{3rdterm1pi}
\end{split} 
\een  
\item{The 4th term is bounded by}
\ben
\begin{split}
\left|\,\frac{n-4}{n(n-1)}\, \partial^{l}_{\mu} \, h_{n}(\mu)\,\right| 
\ \le\ \frac{n-4}{n(n-1)}\ {\cal B}(n,l)\ \le\
\frac{1}{K^2}\ {\cal B}(n+2,l) \ .
\label{4thterm1pi}
\end{split} 
\een  
\end{itemize}
The claim then follows since
\[
\frac{(1+\de^2(\mu))}{c}\left( 
 \frac{\de(\mu)}{2\pi^2}\, \,+\, \frac{(36\,c)^2}{\pi^2}
\,+\, \frac{1}{20\,K}\,+\, \frac{1}{K^2}\right)\le \ 1\ .
\] 
\qed 

\vspace{.2cm}

Proceeding in the same way as we did in proving Proposition \ref{geom3}
as a consequence of Proposition \ref{geom2}, we may deduce from
Proposition  \ref{1pI4} that {\it the smooth solutions we have constructed
are nontrivial and asymptotically free for $\,\beta >0\,$.} 
As stated before we have no result for the 1PI functions so far, assuring
the existence of solutions with bounded action in the sense 
of \ref{ssec:solbounded}. 
In a first moment 
it seems that the boundary conditions for the 1PI functions
$\, {\Ga}^{\al_0,\al}_{n}(p_1,\ldots p_{n})\,$
are easy to analyse because for $\al =\al_0$ 
the $\,C_{\Ga_2}^{\al_0,\al_0}(p)\,$ vanish so that we are only 
left with the contribution $\,v=1\,$ in  \eqref{fega2}. 
But the construction of smooth solutions requires 
control of all derivatives of the $\,h_n(\mu)\,$.
To make further reaching statements this requires more stringent bounds 
on all derivatives of the  $\,J_\nu(\mu)\,$ and of $\,J_0^{-1}(\mu)\,$.

In conclusion we hope that progress will be made in the future on the issues
raised by this paper. Technical improvements should  allow
to control larger values of the couplings, to prove sharper bounds and to take
the limit $\,\al \to \infty\,$ while introducing a finite mass $\,m\,$.
They may also permit to extend all the results for the moments of the
effective action to those of the effective potential. Better control 
might also help to establish a kind of phase diagramme  which characterises
the different types of solutions in their dependence on respective classes of
boundary conditions. It seems natural to us to focus on smooth   
solutions of the FEs if the regulators are chosen to be smooth with respect
to the flow parameter. But this restriction might also deserve further attention.
The most interesting and most challenging problem is certainly to extend our
reasoning beyond the mean field limit.


\newpage

\addcontentsline{toc}{chapter}{Bibliography}

\end{document}